\documentclass[aps,showpacs,prb,twocolumn,superscriptaddress,floatfix,longbibliography]{revtex4-2}
\usepackage{graphicx}
\usepackage{dcolumn}
\usepackage{bm}
\usepackage{color}
\usepackage{comment}
\usepackage{diagbox}
\usepackage{verbatim}
\usepackage{multirow}

\begin{document}


\title{Box model of quantum annealing}

\author{Yang Wei Koh}
\email{patrickkyw@gmail.com}
\affiliation{\relax}

\author{Youjin Deng}
\email{yjdeng@ustc.edu.cn}
\affiliation{Hefei National Research Center for Physical Sciences at the Microscale and School of Physical Sciences, University of Science and Technology of China, Hefei 230026, China}
\affiliation{Hefei National Laboratory, University of Science and Technology of China, Hefei 230088, China}
\affiliation{Shanghai Research Center for Quantum Science and CAS Center for Excellence in Quantum Information and Quantum Physics, University of Science and Technology of China, Shanghai 201315, China}


\date{\today}

\begin{abstract}

A particle-in-a-box model of continuous space quantum annealing is proposed and studied numerically by solving the Schr\"odinger wave equation directly. Three types of energy landscapes with multiple local minima are considered, namely a sinusoidal wave modulated by a concave, a convex, or a flat envelope. Both static (energy spectrum) and dynamical (residual energy) behaviors are analyzed in detail, paying particular attention to the effects of landscape roughness and annealing depth. Simulation results show that the residual energy as a function of annealing speed is largely independent of these two factors. The prevalence of diabatic transitions during annealing is observed, and the discrepancy between our numerical results and the Landau-Zener formula is discussed. An interesting feature in the energy gap spectrum, which we call flat gaps, is examined. Based on it, we propose a mechanism to explain the trapping of wave function in local minima during diabatic transitions, widely observed in our data.

\end{abstract}


\maketitle



\section{Introduction}
\label{sec.introduction}

Quantum annealing (QA) is a metaheuristic for solving optimization problems using quantum fluctuations \cite{Kadowaki98,Farhi01,Santoro02,Das08,Das15,Albash18,Hauke20}, and has predominantly been applied to problems with discrete variables \cite{Young08,Jorg10,Young10,Bapst12,Seki12,Liu15,Mukherjee18}. Recently, QA of continuous variables has been receiving some attention \cite{Finnila94,Stella05,Inack15,Chancellor19,Abel21a,Abel21b,Abel22,Arai23}. An early work in this area by Finnila \emph{et al.} \cite{Finnila94} employed an imaginary-time Schr\"odinger equation approach, which lacks intrinsic quantum effects. Pertinent to our present paper are the studies by Stella \emph{et al.} \cite{Stella05} and by Inack and Pilati \cite{Inack15}, which studied the QA of a multi-valley potential function originally proposed by Shinomoto and Kabashima in the context of simulated annealing (SA) \cite{Shinomoto91}. In Stella \emph{et al.}'s work, the potential is treated phenomenologically in terms of a coarse-grained tight-binding Hamiltonian, and they found a power-law decay of the residual energy with respect to the annealing time. This implies an exponential improvement in performance compared to SA, in which the decay is inverse logarithmic \cite{Shinomoto91}. Inack and Pilati adopted a projective quantum Monte Carlo approach in their simulations, which is imaginary-time in nature, and they also found a power-law decay. On the experimental side, with advances in quantum hardware, some workers have also started using D-Wave machines to perform QA of continuous variables \cite{Chancellor19,Abel21a,Abel21b,Abel22,Arai23}. For instance, Arai \emph{et al.} \cite{Arai23} studied the QA of the Shinomoto-Kabashima potential from this angle, using domain-wall encoding to map the continuous system onto discrete Ising variables. 

A recent work studied the QA of a one-dimensional particle in the following potential \cite{Koh22}
\begin{equation}
V(x)
=
\frac{1}{2}
kx^2
+
\frac{h_0}{2}
\left[
1-
\cos\left(
\frac{2\pi x}{w_0}
\right)
\right]
\label{eq.Vsk(x).definition}
\end{equation}
The first term is a harmonic potential with spring constant $k$. The second term is a cosine wave where $h_0$ and $w_0$ are parameters controlling the amplitude and wavelength, respectively. The total potential $V(x)$ has the form of an undulating wave superimposed upon a quadratic function, simulating a    convex-shaped basin with rugged energy landscape. Equation (\ref{eq.Vsk(x).definition}) was inspired by the work of Shinomoto and Kabashima mentioned ealier. In the field of non-convex continuous-optimization problems, $V(x)$ is also known as the Rastrigin function \cite{Dieterich12}. The function $V(x)$ has a unique global minimum with potential zero located at the origin, and this facilitates the evaluation of any optimization algorithm since the exact solution is known. At the same time, the presence of many local minima makes it a challenging problem for any optimization algorithm to solve. Reference \cite{Koh22} compared the performances of QA and SA in optimizing $V(x)$. It was found that the residual energy, under certain conditions, decays as a power law, in agreement with Refs. \cite{Stella05} and \cite{Inack15}. Other non-classical aspects of QA such as nonlinear schedules and tunneling were also studied in Ref. \cite{Koh22}, exhibiting different novel advantages that QA has over SA. 

Although the functional form of Eq. (1) is already quite simple, there are still many fundamental issues concerning QA in one-dimensional space which are difficult to address using $V(x)$ as a model. One important question is how the number of local minima on the potential surface (i.e., its ruggedness) affects QA. Reference \cite{Koh22} focused on Eq. (\ref{eq.Vsk(x).definition}) for just $h_0=w_0=0.2$. In principle, one can tune these two parameters to adjust the ruggedness of $V(x)$. In practice, however, we found that this led to rapid increase in the number of grid points needed to represent the spatial wave function of the particle. The reason is because one needs to allocate a sufficient number of grid points within every local minimum in order to resolve the wave function accurately. As the number of local minima increases, so does the number of grid points, resulting in higher computational costs. 

Another issue concerns how QA depends on the particle's mass, which plays the role of annealing parameter in continuous space QA. Theoretically, one should tune the mass from zero to infinity continuously in a single trajectory. As mass increases the particle's wavelength decreases, so one must utilize a grid that is wide enough to accommodate the initial delocalized wave function as well as dense enough to resolve its final sharply-peaked waveform. To simultaneously satisfy these two conditions requires a grid with a large number of points. In Ref. \cite{Koh22}, QA was performed in short stages, each spanning a manageable mass range. However, this pragmatic protocol may not necessarily yield the same insights as annealing over a single long stage spanning a wider mass range. There is therefore a need to study QA under conditions which are closer to the original theoretical prescription.

One way to circumvent the limitations mentioned above is to use the energy basis of the harmonic oscillator instead of the $x$-representation. In the energy representation, the integration over $x$ can be carried out analytically, avoiding the need to use a space grid. While this is certainly a feasible approach when considering energy-related quantities, it becomes somewhat awkward when we wish to visualize the wave function in the $x$-representation. This is because the basis functions of the harmonic oscillator representation involves Hermite polynomials $H_n(x)$, which for large $n$ requires evaluation of large factorials. We found that for $n\gtrsim 300$, standard numerical packages cease to return values for $H_n(x)$ due to numerical overflows. While this difficulty is not necessarily an unsolvable problem, we would like to take another approach by adopting a different model.

\begin{figure}[h]
\begin{center}
\includegraphics[scale=0.60]{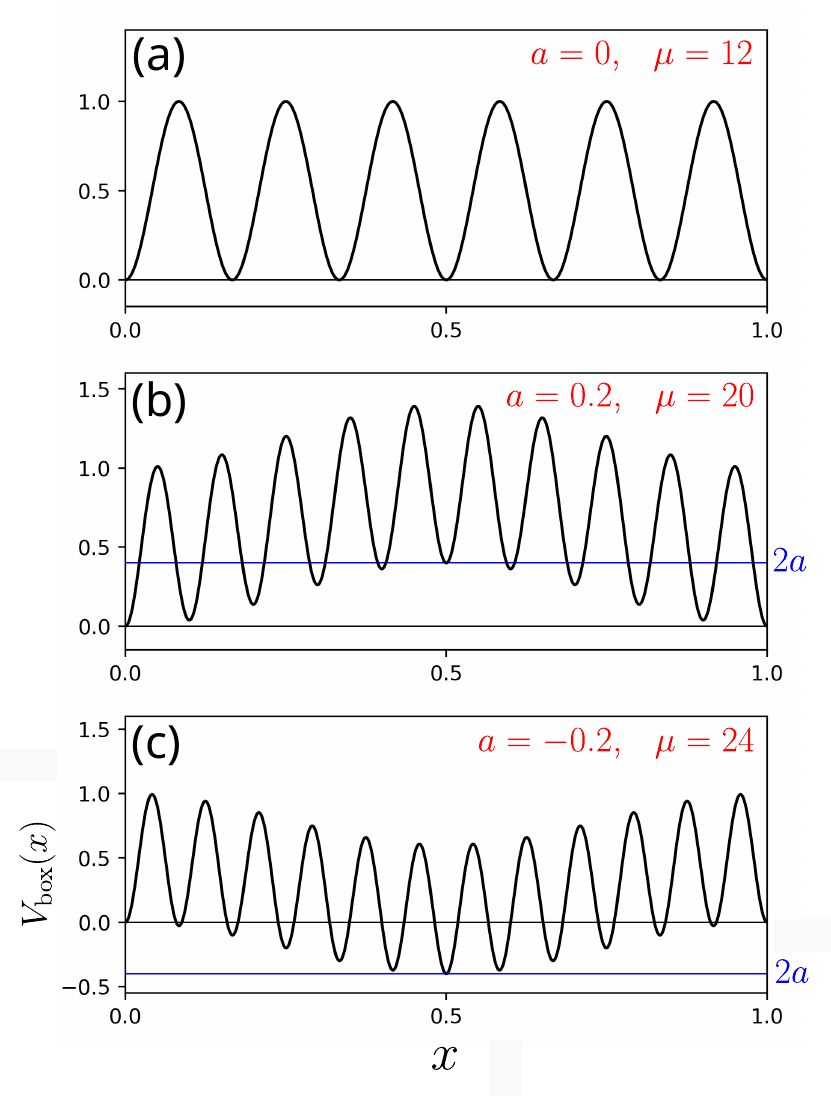}
\caption{The three types of potential surfaces that can be generated by $V_{\mathrm{box}}(x)$ [Eq. (\ref{eq.Vbox.definition})]. The parameter $\mu$ controls the number of minima on the surface. (a) When $a=0$. All minima are ground states, with zero potential. (b) When $a>0$. The surface has a concave envelope, and the minima with lowest potential are located at the walls. (c) When $a<0$. The surface has a convex envelope. The global minimum is located at $x=0.5$. The horizontal line indicated as $2a$ (blue) in panels (b) and (c) shows the vertical displacement of the center minimum.}
\label{fig.3 types V(x)}
\end{center}
\end{figure}

Consider a one-dimensional particle in a box with a finite width. Let the potential outside the box be infinitely high, and the potential inside be $V_{\mathrm{box}}(x)$, which serves as our cost function. We can use the energy basis of the free particle to represent the system for arbitrary $V_{\mathrm{box}}(x)$. Since the bases are just trigonometric functions, the problem of numerical overflow discussed above does not happen here. Deferring the formal definition of $V_{\mathrm{box}}(x)$ to Sec. \ref{sec.model}, Fig. \ref{fig.3 types V(x)} shows some examples of potential surfaces that we shall study in this work. It is seen that restricting the particle to a box gives us more flexibility to design the cost function. For instance, panel (a) shows a surface with many degenerate minima, while panel (b) shows a concave-shaped surface. Such landscapes would be rather cumbersome to define and to simulate if the particle is allowed to move on the entire $x$-axis. At the same time, one can also generate surfaces similar to the Rastrigin function, as shown in panel (c). Another advantage of our box model is that the number of local minima on the landscape can be increased without worrying about allocating an appropriate spatial grid each time, greatly facilitating the study of how landscape coarseness affects QA performance. Thirdly, working in the kinetic energy representation, we can now anneal from a small to a large mass directly in a single trajectory, thereby avoiding the artificially-defined annealing stages utilized in Ref. \cite{Koh22}. These are fundamental questions concerning continuous space QA that are unfeasible to address using Eq. (\ref{eq.Vsk(x).definition}) as a model, at least from the standpoint of numerical simulations.

The rest of the paper is organized as follows. The model is defined in Sec. \ref{sec.model}. Sections \ref{sec.a=0}, \ref{sec.a=0.2}, and \ref{sec.a=-0.2} are devoted to the three classes of energy landscapes shown in panels (a), (b), and (c) of Fig. \ref{fig.3 types V(x)}, respectively. In each section, the discussion proceeds in two parts. We first consider the static behavior of the system, particularly the energy levels and the eigenfunctions of the ground and low-lying states. The second part presents the results of annealing. We assessed the efficiency of QA by looking at the rate at which residual energy decays as a function of annealing time. The dependences of the decay rate on landscape roughness and annealing depth (i.e., the initial and final masses) are examined in detail. Section \ref{sec.conclusion} summarizes and concludes the paper. To improve the flow of our discussion, some materials are deferred to the appendices. Appendix \ref{app.R(T).a=0.adiabatic.derivation} derives the adiabatic approximation of the residual energy. Appendix \ref{app.variational method to flat gap} analyzes the energy gap using the variational method. The reader is referred to them for technical details.


\section{Model}
\label{sec.model}

\subsection{Potential energy}
\label{}

As mentioned, we consider a one-dimensional particle in the region $0<x<L$. The potential energy in this region is defined as  
\begin{equation}
V_{\mathrm{box}}(x)
=
V_{\mu}(x) + V_{a}(x)
\label{eq.Vbox.definition}
\end{equation}
and consists of two terms. The first term $V_{\mu}(x)$ is
\begin{equation}
V_{\mu}(x)=
\frac{1}{2}
\left[
1-
\cos
\left(
\frac{\mu\pi x}{L}
\right)
\right]
\label{}
\end{equation}
where the parameter $\mu$ determines the number of minima on the potential surface. For our purposes, we only consider $\mu$ which are positive integers and multiples of four (i.e. $\mu=4,8,\cdots$). The second term $V_{a}(x)$ is
\begin{equation}
V_{a}(x)=
a
\left[
1-
\cos
\left(
\frac{2\pi x}{L}
\right)
\right]
\label{eq.Va(x).definition}
\end{equation}
where the parameter $a$ is a real number controlling the overall envelope of the potential surface. When $a=0$, $V_{\mathrm{box}}(x)$ is just a simple plane wave. When $a>0$ ($a<0$), the wave is modulated by a concave (convex) envelope. Without loss of generality, in numerical simulations we let the box width $L$ be 1.

Figure \ref{fig.3 types V(x)} illustrates three instances of potential surface generated by $V_{\mathrm{box}}(x)$. Panel (a) shows the case of flat envelope $a=0$, with $\mu=12$. It is seen that all the energy minima are degenerate with zero potential. Panel (b) shows the case of concave envelope $a=0.2$ with $\mu=20$. In this case, the minima with the lowest potential are located at the walls $x=0$ and $x=L$. The center minimum at $x=\frac{L}{2}$ has potential $2a$, as indicated by the horizontal line (blue). Panel (c) shows a convex case with $a=-0.2$ and $\mu=24$. The center minimum is now the (unique) global minimum with potential $2a$. From the figure, it is straightforward to verify that the total number of minima on the potential surface is $\frac{\mu}{2}+1$ (including the two at the walls $x=0, L$).

There are two adjustable parameters in Eq. (\ref{eq.Vbox.definition}), $\mu$ and $a$. In our numerical calculations, we shall mainly focus on varying $\mu$, which controls the number of minima on $V_{\mathrm{box}}(x)$. For the envelope amplitude $a$, we limit ourselves to just three values 0, 0.2, and $-0.2$, which we feel is representative of the three classes of envelopes discussed above.

Concerning nomenclature, the potential surfaces for $a$ positive and negative are, mathematically speaking, not concave and convex. Nevertheless, in the context of this paper, we shall sometimes simply refer to them as concave and convex systems, for a more fluent discussion.

\subsection{Annealing Hamiltonian}
\label{}

Let us now introduce the annealing Hamiltonian,
\begin{equation}
H(s)=\frac{1}{s} \left( \frac{p^2}{2m} \right) + V_{\mathrm{box}}(x)
\label{eq.H(s).definition}
\end{equation}
where $p$ is the momentum operator $\frac{\hbar}{i}\frac{d}{dx}$, $m$ is the mass of the particle, and $s$ is a positive real number which serves as the annealing parameter. Without loss of generality, we let $m$ and $\hbar$ be 1 in numerical calculations.

In quantum annealing parlance, $V_{\mathrm{box}}(x)$ plays the role of problem Hamiltonian, the kinetic energy operator $\frac{p^2}{2m}$ serves as the driver, and $s$ controls the relative strength between these two terms. One begins annealing with a small $s$ such that the kinetic term is dominant. The input wave function (i.e., kinetic energy ground state) is delocalized in coordinate space, allowing the particle to fully explore the landscape of $V_{\mathrm{box}}(x)$ in search of the global minimum. One then gradually increases $s$, which effectively makes the particle more massive, slowing it down. Annealing is terminated when $s$ reaches a certain large value, at which point the kinetic energy is almost completely suppressed and $V_{\mathrm{box}}(x)$ becomes the dominant term in the Hamiltonian. The particle's wave function is now spatially localized, and (hopefully) located at the global minimum of $V_{\mathrm{box}}(x)$. 

We briefly comment on the form of our annealing Hamiltonian Eq. (\ref{eq.H(s).definition}) vis-\`a-vis the more familiar form based on convex combination
\begin{equation}
\tilde{H}(\tau)
=
(1-\tau)
\left( \frac{p^2}{2m} \right)
+
\tau
\,
V_{\mathrm{box}}(x)
\label{eq.H(tau)}
\end{equation}
where the annealing parameter $\tau$ increases from 0 to 1. We experimented with Eq. (\ref{eq.H(tau)}) and found that the ground state wave function of $\tilde{H}(\tau)$ exhibits very little change unless $\tau$ approaches extremely close to 1. This somewhat `lopsided' dependence on the annealing parameter (of the Hamiltonian's properties) makes Eq. (\ref{eq.H(tau)}) unwieldy to study, which is why we prefer Eq. (\ref{eq.H(s).definition}) instead.

\subsection{Kinetic energy representation of Hamiltonian matrix}
\label{}

To represent the Hamiltonian Eq. (\ref{eq.H(s).definition}), we use the basis of the kinetic energy operator $\frac{p^2}{2m}$. Its eigenvalues are 
\begin{equation}
K_n=\frac{(n+1)^2 \pi^2 \hbar^2}{2mL^2}
\label{eq.Kn.definition}
\end{equation}
with the corresponding eigenfunctions
\begin{equation}
\phi_n(x)
=
\sqrt{\frac{2}{L}}
\sin
\left[
\frac{(n+1)\pi x}{L}
\right]
\label{eq.phin(x).definition}
\end{equation}
where $n$ ($=0,1,2,\cdots$) is the basis label. In this representation, the first term in Eq. (\ref{eq.H(s).definition}) is a diagonal matrix with elements given by Eq. (\ref{eq.Kn.definition}). The second term $V_{\mathrm{box}}(x)$ contains off-diagonal matrix elements due to the cosine functions, and they are given by
\begin{eqnarray}
&& \int_0^L
\phi_n(x)
\phi_m(x)
\cos
\left(
\frac{l \pi x}{L}
\right)
dx \nonumber \\
&& =
\frac{1}{2}
\left(
\delta_{n-m+l}
+
\delta_{n-m-l}
-
\delta_{n+m+2-l}
\right)
\nonumber 
\end{eqnarray}
where $n,m \ge 0$ and $l\ge 2$ are integers, and the delta symbol $\delta_p$ is defined as 
\begin{equation}
\delta_p=
\left\{
\begin{array}{ccc}
1& \mathrm{if}& p=0,\\
0& \mathrm{if}& p\ne 0\\
\end{array}
\right.
\label{}
\end{equation}

A wave function in the $x$-representation $\psi(x)$ is related to the vector elements of its kinetic energy representation $c_n$ as
\begin{equation}
\psi(x)=\sum_{n=0}^{\infty} c_n \phi_n(x)
\label{eq.psi(x)=cn}
\end{equation}
In actual numerical calculations, the upper limit in Eq. (\ref{eq.psi(x)=cn}) must be truncated at a finite number, which we denote as $N_{\mathrm{dim}}$ from now on.


\section{Potential surface with many degenerate minima (\lowercase{\emph{a}}=0)}
\label{sec.a=0}

For our first example, we consider the case where the envelope of $V_{\mathrm{box}}(x)$ is flat, obtained by setting $a=0$ in Eq. (\ref{eq.Va(x).definition}). The potential surface when $\mu=12$ is shown in Fig. \ref{fig.3 types V(x)}(a). The surface is a cosine wave, and all the minima are degenerate with zero potential. Although simple, one might consider such a potential as emulating certain aspects of spin-glasses. It is well-known that the energy landscapes of spin glasses possess many degenerate ground states. The important difference here is that on our surface there are no local minima acting as metastable traps. Another difference is that our ground states are all exactly degenerate rather than quasi-degenerate. Thirdly, our local minima are evenly distributed in configuration space, whereas in a spin glass the ground states' distribution is random and their precise locations generally unknown. Despite these differences, it is nevertheless interesting to see how QA will perform when faced with a simplified and somewhat idealized system.

\subsection{Energy levels and the ground state}
\label{subsec.a=0.En and gds}

\begin{figure}[h]
\begin{center}
\includegraphics[scale=0.55]{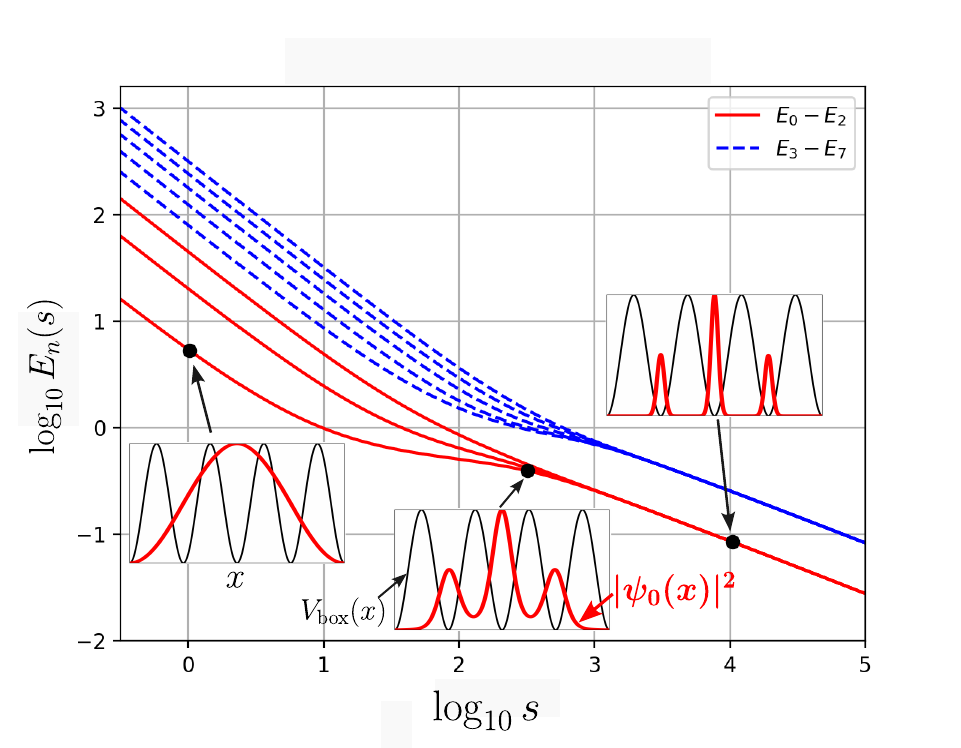}
\caption{Energy levels $E_{0-7}(s)$ of the $a=0$, $\mu=8$ system ($N_{\mathrm{dim}}=1000$). There are five degenerate minima on the potential surface. Solid lines (red) show $E_{0-2}$, while dashed lines (blue) show $E_{3-7}$. It is seen that $E_{0-2}$ merge into a single level at $\log_{10}s\approx 2.5$, due to the three center minima. Insets show the probability densities of the ground state $|\psi_0(x)|^2$, superposed on $V_{\mathrm{box}}(x)$, at values of $s$ indicated by solid circles.}
\label{fig.Ens.a=0 and mu=8}
\end{center}
\end{figure}

Let us first examine the static behavior of the system. Figure \ref{fig.Ens.a=0 and mu=8} shows the energy levels when $\mu=8$, obtained by  diagonalizing the Hamiltonian $H(s)$ in the kinetic energy representation numerically. The eight lowest levels are plotted as a function of the annealing parameter $s$ (note the logarithmic scales). The solid curves (red) show $E_0$ to $E_2$, while the dashed ones (blue) show $E_3$ to $E_7$. It is seen that $E_{0-2}$ merge into a three-fold degenerate level at $\log_{10}s\approx 2.5$. This is due to three center minima on the surface of $V_{\mathrm{box}}(x)$, which can be understood by looking at the ground state wave function $\psi_0(x)$. The insets show the probability density $|\psi_0(x)|^2$ at some representative values of $s$, indicated by solid circles. The graphs of $|\psi_0(x)|^2$ are superposed on that of $V_{\mathrm{box}}(x)$, which has three center minima and two minima at the walls. When $s=1$, the ground state density is a broad, gauss-like curve straddling the three central minima. As $s$ increases, the density evolves from having a single wide hump to having three narrow peaks each localized within the central minima, as can be seen from the insets of $\log s=2.5$ and 4. It is straightforward to generalize this observation to the case of arbitrary $\mu$. When there are $n_{\mu}=(\frac{\mu}{2}-1)$ central degenerate minima on the surface of $V_{\mathrm{box}}(x)$, as $s$ increases from some small value, the $n_{\mu}$ lowest energy levels will eventually merge into a single $n_{\mu}$-fold degenerate ground state. The degenerate level is spanned by $n_{\mu}$ independent gaussian peaks, each localized in one of the $n_{\mu}$ degenerate minima.

It is interesting to note that even though from a classical point of view all the central minima of $V_{\mathrm{box}}(x)$ are energetically equivalent, from a quantum perspective they are not. This can be evinced from the ground state density at $\log s=2.5$ shown in the middle inset. We see that the peak at $x=\frac{1}{2}$ is higher than the other two at $x=\frac{1}{4}$ and $\frac{3}{4}$. This preference for the center minimum can be understood as remnants of the ground state's profile from an earlier $s$, which is peaked at the center (e.g. at $s=1$). Another way to look at  this is that each minimum of $V_{\mathrm{box}}(x)$ is different in view of its distances from the two walls, and this information is captured by the ground state wave function. This is a manifestation of the wholeness of a quantum system, and is utilized during QA. We believe this to be an important feature which is not present in classical optimization algorithms such as simulated annealing. In classical physics, only the energy of the minimum matters during optimization, not its relationship to the rest of the energy landscape. On the other hand, based on this we can also anticipate that during actual annealing, the off-centered minima are less likely to be found. In other words, minima with the same energy may not be treated equally in QA.

Another interesting observation is that the two minima at the walls are avoided by the ground state wave function, even though the potential energy at these two points are the same as the other minima. This is a consequence of boundary conditions, which require that the wave function vanishes at the walls. The region in the vicinity of a wall can be approximated by a \emph{half} harmonic oscillator \cite{Griffths94}, which can be solved exactly like the harmonic oscillator. Its ground state energy is $\frac{3}{2}\hbar\omega$, which is one quantum more than that of a normal oscillator whose angular frequency is $\omega$. This means that the states at the walls have higher energy than those at the other minima, so they can never appear inside the ground state's degenerate energy level. Indeed, in our numerical calculations these `wall states' are found in the next degenerate level (i.e., the blue dashed level in Fig. \ref{fig.Ens.a=0 and mu=8}).

\begin{figure}[h]
\begin{center}
\includegraphics[scale=0.55]{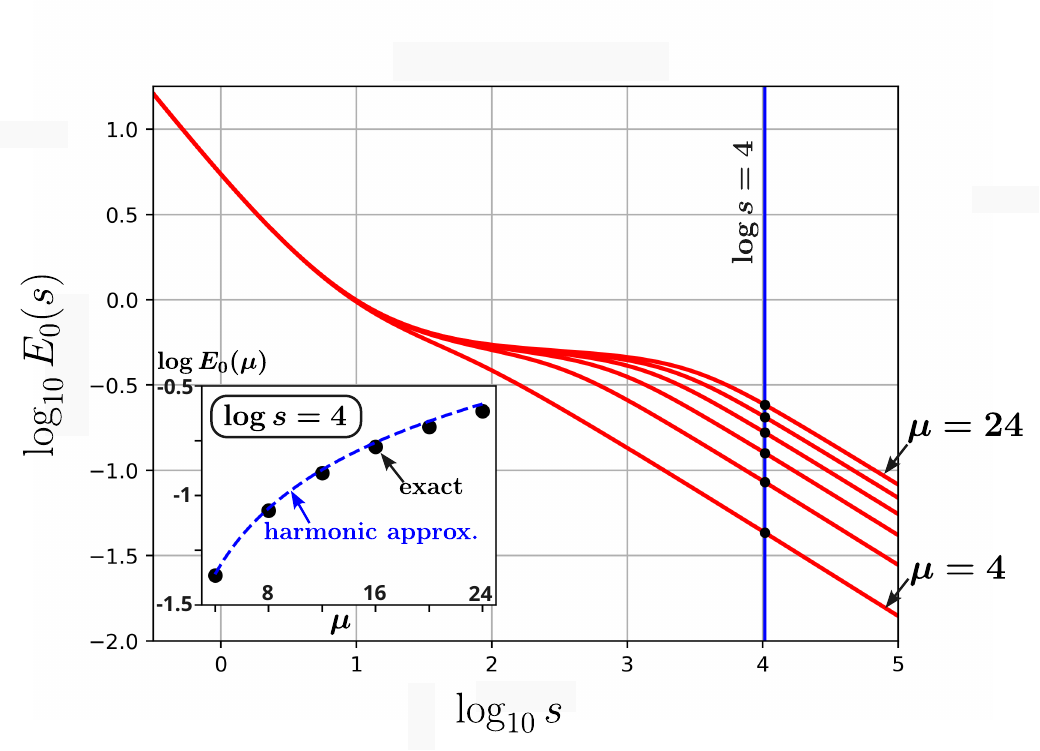}
\caption{Ground state energy $E_0(s)$ of the $a=0$ system as $\mu$ increases from 4 to 24. When $s$ is small kinetic energy dominates, so all curves fall into one, independent of $V_{\mathrm{box}}(x)$. As $s$ increases, the curves split, and $E_0$ increases as $\mu$ increases. This is highlighted by the vertical line (blue) at $\log s=4$. The solid circles (black) are replotted in the inset, exhibiting the dependence of $E_0$ on $\mu$. The dashed curve (blue) shows our theoretical prediction based on the zero-point energy of a harmonic oscillator, given by Eq. (\ref{eq.a=0.ZPE.harmonic approx}).}
\label{fig.ZPE.a=0.various mu}
\end{center}
\end{figure}

Let us now look at how the number of minima affects the ground state energy. Figure \ref{fig.ZPE.a=0.various mu} shows the graphs of $E_0(s)$ for several values of $\mu$. When $s$ is small, kinetic energy dominates, so the graphs of different $\mu$ merge, independent of $V_{\mathrm{box}}(x)$. As $s$ increases, the curves split, with those having larger $\mu$ attaining higher energies. That the ground state energy should increase with $\mu$ might be somewhat counter-intuitive from a classical standpoint, since all the minima are at zero potential. Nevertheless, this is again a consequence of quantum behavior, which requires us to take zero-point energy into account; specifically, the increase in curvature of the local minima as $\mu$ increases. To be concrete, consider the vertical line at $\log s=4$ in Fig. \ref{fig.ZPE.a=0.various mu}. The increase of $E_0$ with $\mu$ is replotted in the inset (solid circles). If we Taylor expand $V_{\mathrm{box}}(x)$ about any of the minima, one finds the spring constant $\frac{1}{2}\left(\frac{\pi\mu}{L}\right)^2$, yielding the zero-point energy
\begin{equation}
E_{\mathrm{z.p.}}(\mu)
=
\frac{\hbar \pi}{2\sqrt{2s}L}
\,
\mu
\label{eq.a=0.ZPE.harmonic approx}
\end{equation}
The graph of Eq. (\ref{eq.a=0.ZPE.harmonic approx}) is shown as the dashed curve in the inset. We see that our theoretical explanation agrees very well with the observed data.

\subsection{Residual energy under linear annealing schedule}
\label{subsec.annealing.a=0}

We now study the dynamics of the system by solving the time-dependent Schr\"odinger equation numerically in the kinetic energy representation \cite{notes.Scipy.solveivp}. The Hamiltonian Eq. (\ref{eq.H(s).definition}) now depends on time $t$ via $s(t)$ according to the following linear annealing schedule
\begin{equation}
s(t)=(s_f-s_i)\left( \frac{t}{T} \right) + s_i
\label{eq.linear schedule.definition}
\end{equation}
where $T$ is the total annealing time, and the Schr\"odinger equation is integrated from $t=0$ to $t=T$. Equation (\ref{eq.linear schedule.definition}) is termed linear because $s$ depends linearly on $t$. The annealing time $T$ controls the speed of annealing, and should (in principle) be large so that the system evolves slowly. The parameters $s_i=s(0)$ and $s_f=s(T)$ are the initial and final values of $s$. They are chosen by consulting the system's energy levels $E_n(s)$, taking care to include salient features (e.g., phase transitions) so that our annealing path would not be a trivial one. The initial wave function $c_n(t=0)$ [see Eq. (\ref{eq.psi(x)=cn})] is taken to be the ground state of the initial Hamiltonian $H(s_i)$. 

To assess the results of annealing, we consider the residual energy, defined as 
\begin{equation}
R(T)=\left\langle H(s_f) \right\rangle_{T} - E_{\mathrm{ref.}}
\label{eq.Residual Energy.definition}
\end{equation}
The first term is the expectation value of the final Hamiltonian taken with respect to the annealed wave function $c_n(t=T)$. The second term $E_{\mathrm{ref.}}$ is the energy of the target state one is trying to attain. In most cases, $E_{\mathrm{ref.}}$ is taken to be $E_0(s_f)$, the ground state energy of $H(s_f)$. However, it sometimes helps to define it at certain excited states, giving us flexibility in handling situations involving first-order transitions. Equation (\ref{eq.Residual Energy.definition}) is a measure of how quickly QA is able to attain the global minimum of $V_{\mathrm{box}}(x)$ (or the targeted state). If $R(T)$ decays to zero very quickly with respect to $T$, then QA is an efficient algorithm. On the other hand, if $R(T)$ decreases very slowly or stagnates at some positive value, then it is indication that QA is ineffective or has failed.

\begin{figure}[h]
\begin{center}
\includegraphics[scale=0.55]{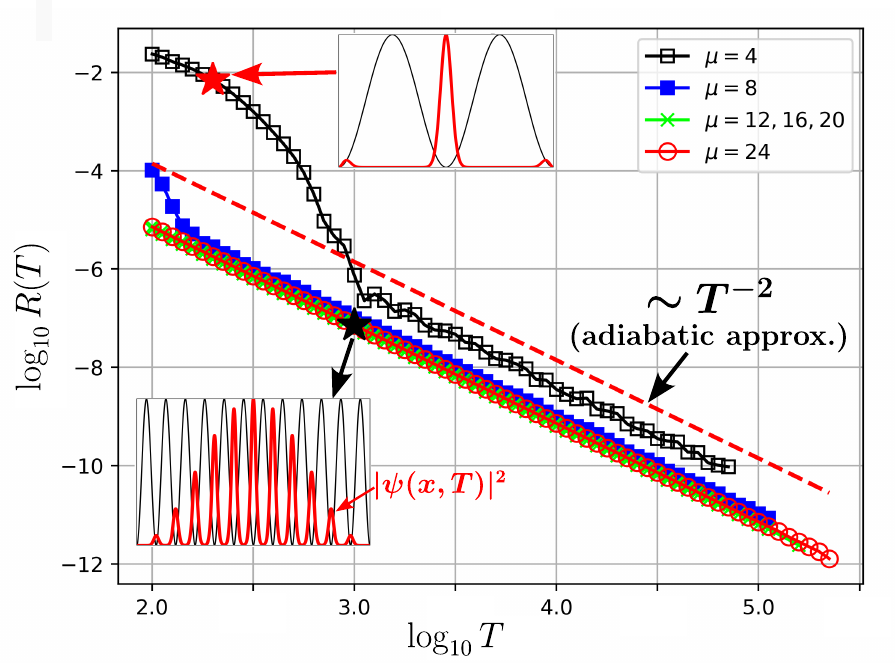}
\caption{Residual energy $R(T)$ of the $a=0$ system, obtained under the linear schedule Eq. (\ref{eq.linear schedule.definition}). The graphs of $\mu=12$ to 24 have collapsed onto one. Dashed line (red) shows $\frac{1}{T^2}\frac{25\sqrt{2}}{8\pi}$, the upper bound on adiabatic approximation Eq. (\ref{eq.R(T).a=0.adaibatic approx}). Insets show the probability densities of the final annealed wave function $|\psi(x,T)|^2$. Lower inset: For $\mu=24$ and $T=1000$. Upper inset: For $\mu=4$ and $T=200$. The two stars indicate the residual energies of the wave functions shown in the insets. In this figure, $s_i=1$, $s_f=10^4$, $N_{\mathrm{dim}}=400$, and $E_{\mathrm{ref.}}=E_0(s_f)$.}
\label{fig.RE curve.a=0}
\end{center}
\end{figure}

Figure \ref{fig.RE curve.a=0} shows the $R(T)$ curves for $\mu$ from 4 to 24. The parameters are $E_{\mathrm{ref.}}=E_0(s_f)$, and $s_i=1$, $s_f=10^4$ for all $\mu$'s. The choices of $s_i$ and $s_f$ were made based on Fig. \ref{fig.ZPE.a=0.various mu}, where we see that the curved portion of all the $E_0(s)$ curves fall within the annealing range $\log s \in [0,4]$. (Note that our annealings span four orders of magnitude of $s$, which is wider than those in Ref. \cite{Koh22}.) The graphs show that, in general, $R(T)$ decays as $T^{-2}$. This scaling is a familiar result, and is indication that the annealing is in the adiabatic regime \cite{Ballentine98,Morita07}. For the present system (i.e., $a=0$), we also derived the following expression for the residual energy under adiabatic approximation (see Appendix \ref{app.R(T).a=0.adiabatic.derivation} for details)
\begin{equation}
R_{\mathrm{adia.}}(T)
=
\frac{\tilde{a}}{T^2}
\left(
1-\frac{s_i}{s_f}
\right)^2
\,
\sin^2(\omega_{f} T)
\,
\left(
1-\frac{2}{\mu}
\right)
\label{eq.R(T).a=0.adaibatic approx}
\end{equation}
where $\omega_f$ is given by Eq. (\ref{eq.app.omegaf.definition}) and $\tilde{a}=\frac{\hbar L \sqrt{2ms_f}}{32\pi}$.
The $T^{-2}$ scaling originates from the term $\frac{dH[s(t)]}{dt}$ which appears when one makes adiabatic approximation. Another point about Eq. (\ref{eq.R(T).a=0.adaibatic approx}) is that the factor $(1-\frac{2}{\mu})$ means that for large $\mu$ the residual energy will be very much independent of $\mu$, which explains why the graphs from $\mu=12$ to 24 seem to have collapsed onto one. This $\mu$-independence is due to cancellation of the increased curvature of each minimum with the total number of minima, as seen from the derivation in Appendix \ref{app.R(T).a=0.adiabatic.derivation}. Approximating the last three factors in Eq. (\ref{eq.R(T).a=0.adaibatic approx}) by unity, the right hand side becomes $\frac{1}{T^2}\frac{25\sqrt{2}}{8\pi}$, which is plotted in Fig. \ref{fig.RE curve.a=0} as the dashed line. One sees that this upper bound on the residual energy agrees very well with the numerical data.

The lower inset in Fig. \ref{fig.RE curve.a=0} shows the probability density of the annealed wave function $|\psi(x,T=1000)|^2$ for $\mu=24$. It is seen that the density is localized in all the energy basins (except at the walls), modulated by a gaussian-like envelope. This corroborates our earlier comment that during annealing not all degenerate global minima would be found with equal probability, due to the form of the initial wave function. 

The graph of $\mu=4$ is slightly different from the rest, exhibiting an exponential decay before crossing over to the adiabatic regime. This is a signature of non-adiabatic transitions \cite{Morita07}, which in simple scenarios can be described by Landau-Zener theory \cite{Landau32,Zener32}. Deferring a detailed discussion to later sections where diabatic transitions become the dominant feature, here we would just like to comment on why only $\mu=4$ exhibits this behavior. For $\mu=4$, the basins at the walls are quite large and overlap substantially with the initial wave function. (This overlap can be discerned from the $s=1$ inset in Fig. \ref{fig.Ens.a=0 and mu=8}, replacing the graph of $V_{\mathrm{box}}(x)$ for $\mu=8$ by $\mu=4$.) Hence, the particle may fall into the basins at the walls and get trapped there. As the wall minima have higher energy than the center minimum, this gives rise to the diabatic regime on the $\mu=4$ curve. The upper inset in Fig. \ref{fig.RE curve.a=0} shows the annealed probability density for $T=200$ on the $\mu=4$ graph. It is seen that the density has two small lobes at the walls. By contrast, for $\mu\ge8$ the wall basins have almost no overlap with the initial wave function (e.g., see $s=1$ inset in Fig. \ref{fig.Ens.a=0 and mu=8}), so there is very little chance for the particle to get trapped there.


\section{Potential surface with concave envelope (\lowercase{\emph{a}}=0.2)}
\label{sec.a=0.2}

We now proceed to the case of concave envelope, obtained by setting $a=0.2$ in Eq. (\ref{eq.Va(x).definition}). The potential surface when $\mu=20$ is shown in Fig. \ref{fig.3 types V(x)}(b). One sees that there are two global minima located at the walls $x=0$ and $L$. This potential is more challenging than the previous one to minimize because there are many local minima. One interpretation of this surface is that the concave envelope represents a large potential barrier separating two deep energy basins, while the local minima represent the basins' rugged interiors. In this system, the particle encounters energy barriers with different orders of magnitude during the course of annealing: high barriers across different basins, as well as smaller ones within individual basins. Such scenarios arise, for instance, in the QA of the Hopfield model \cite{Knysh16,Koh18}, where the deep basins correspond to different memory states stored in the neural network. Although energy landscapes like this occur frequently in many-body systems, their QA are quite difficult to simulate numerically due to exponentially large Hilbert spaces. Here, our model offers a simplified one-dimensional caricature of the problem which can be studied in considerable detail.

\subsection{Energy gap closure and first-order transition}
\label{subsec. a=0.2 .gap closure and 1st order}

\begin{figure}[h]
\begin{center}
\includegraphics[scale=0.50]{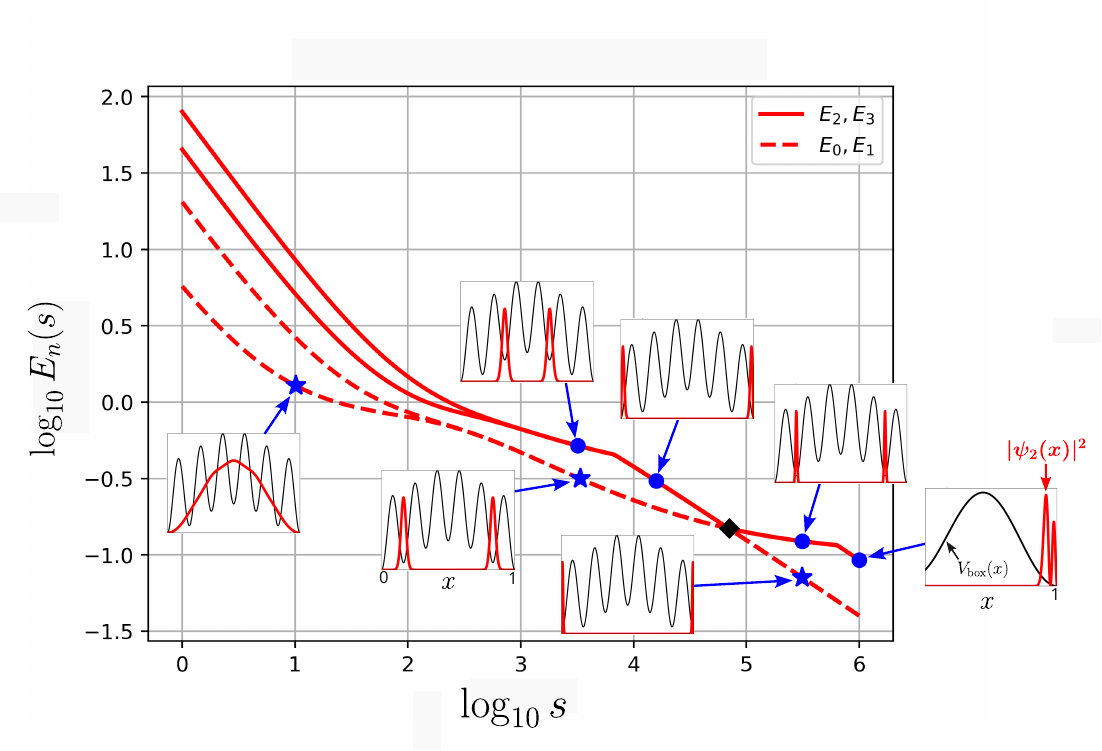}
\caption{Energy levels $E_{0-3}(s)$ of the $a=0.2$, $\mu=12$ system ($N_{\mathrm{dim}}=1000$). The potential $V_{\mathrm{box}}(x)$ has a concave envelope and seven minima. As $s$ increases, $E_{0,1}$ (dashed lines) merge into a doubly-degenerate ground state, while $E_{2,3}$ (solid lines) into a doubly-degenerate first-excited state. At $\log s\approx 4.85$ (black diamond) the gap between the ground and first-excited states vanishes. Insets show the probability densities of the energy eigenfunctions $|\psi_n(x)|^2$ at some representative values of $s$, indicated by the stars (for $n=0$) and solid circles ($n=2$). This figure is a roadmap, providing an intuitive sense of what to expect during finite-time QA.}
\label{fig.E0123.a=0.2 and mu=12}
\end{center}
\end{figure}

As before, we first look at the statics of the system. Figure \ref{fig.E0123.a=0.2 and mu=12} shows the four lowest energy levels for the case of $\mu=12$, where the potential surface has seven minima (including the two at the walls). The dashed lines show the graphs of $E_0$ and $E_1$ as a function of $s$, while the solid lines show those of $E_2$ and $E_3$. As $s$ increases, $E_{0,1}$ merge into a doubly-degenerate ground state at $\log s\approx 2$, and $E_{2,3}$ merge to become the first-excited state at $\log s\approx 2.5$. At $\log s\approx 4.85$ (black diamond), the energy gap closes and the ground state is four-fold degenerate at that point. There are no other gap closures with further increase in $s$. The overall features of this energy level diagram for $\mu=12$ is representative of other $\mu$ as well, with only minor differences. One particular point to note is that as $\mu$ increases, the gap closes at larger $s$.

The insets show the probability densities of the energy eigenstates $|\psi_n(x)|^2$ at selected values of $s$, which are indicated by stars (for $n=0$) and circles (for $n=2$). As $s$ increases, the ground state density evolves from a broad curve ($\log s=1$) into two peaks centered at the first and last local minima ($\log s\approx 3.5$), and finally into two peaks located at the walls ($\log s\approx 5.5$). To facilitate our future discussions, for any $\mu$ in general, let us call the two local minima which are closest to the walls `adjacent minima'. At the point where the energy gap closes, the ground state density undergoes a first-order transition and jumps from the adjacent minima to the global minima. It is known that first-order transitions are very hard for QA to overcome \cite{Young08,Jorg10,Young10}. Consequently, even though the global minima at the walls are in principle the true ground states of the system (at sufficiently large $s$), in practice one can only expect to minimize until the adjacent minima during annealing. We also mentioned earlier that as $\mu$ increases, the gap closure migrates towards larger $s$. Hence, generally for the concave system, we should consider the adjacent minima as the lowest energy states that can be attained during QA, and set $E_{\mathrm{ref.}}$ in Eq. (\ref{eq.Residual Energy.definition}) to be the energy of these states.

The upper insets indicated by circles show the density $|\psi_2(x)|^2$ at some values of $s$. It is seen that the first-excited state also has the form of gaussian peaks centered at local minima. As $s$ increases, the peaks jump from the second lowest minima ($\log s\approx 3.5$) to the global minima, then to the adjacent minima, and finally back to the global minima ($\log s=6$). The first and third jumps are due to avoided crossings with the second-excited state (not shown in the figure). The fact that the first-excited state has the form of a gaussian is useful information because we can apply the variation method with a gaussian ansatz to derive the energy of the first-excited state, and hence the energy gap of the system. Such a semi-classical approach to the energy gap is usually not applicable when dealing with many-body spin systems. We shall revisit this when we discuss the convex system in Sec. \ref{sec.a=-0.2}.

As a final point, note that at $\log s=6$ the first-excited state is inside the global minima and exhibits a single node, which is in accordance with our usual understanding of its nodal structure. The transition of the first-excited state into the global minimum can be used as an indication that the kinetic energy has become sufficiently weak such that the system is in the classical regime.


\subsection{Zero-point energies at the global and adjacent minima}
\label{}

We now examine the energy states more quantitatively.

The first-order transition discussed above can be understood as competition between quantum and classical energies. It was mentioned in Sec. \ref{subsec.a=0.En and gds} that the zero-point energy of a half oscillator is one quantum more than that of a normal oscillator. The energy at the wall minimum is therefore one quantum higher than at the adjacent minimum. On the other hand, the potential of the adjacent minimum is higher than the wall minimum. Hence, the transition point (i.e., gap closure) occurs when the additional quantum and the extra potential energy balance each other.

Repeating the same steps as in the derivation of Eq. (\ref{eq.a=0.ZPE.harmonic approx}), the zero-point energy at the wall minimum is  
\begin{equation}
E_{\mathrm{wall}}(\mu)
=
3
\cdot
E_{\mathrm{z.p.}}(\mu)
\cdot
\sqrt{1+\frac{8a}{\mu^2}}
\label{eq.a=0.2.zpe of half oscillator}
\end{equation}
where the factor of 3 takes into account the  additional quantum, the radical containing $a$ is contributed by the curvature of the concave envelope, and $E_{\mathrm{z.p.}}(\mu)$ is given by Eq. (\ref{eq.a=0.ZPE.harmonic approx}). For the potential energy, the $x$-coordinate of the adjacent minimum $x_{\mathrm{adj.}}$ is approximately 
\begin{equation}
x_{\mathrm{adj.}}=\frac{2L}{\mu}
\label{}
\end{equation}
which gives an excess potential energy of 
\begin{equation}
V_{\mathrm{box}}(x_{\mathrm{adj.}})
=
a\left[
1-\cos\left(\frac{4\pi}{\mu}\right)
\right]
\label{eq. a=0.2. Vx1(mu)}
\end{equation}
relative to the wall minimum. Equating Eq. (\ref{eq. a=0.2. Vx1(mu)}) with the additional quantum in Eq. (\ref{eq.a=0.2.zpe of half oscillator}), one obtains the point of first-order transition 
\begin{equation}
s_{\mathrm{1st}}
=
\left(\frac{\pi\hbar}{aL\sqrt{2}}\right)^2
\frac{\mu^2+8a}{\left[1-\cos\left(4\pi/\mu\right)\right]^2}
\label{eq.s.1st order}
\end{equation}
From Eq. (\ref{eq.s.1st order}) one readily verifies that for $a=0.2$ and $\mu=12$ one has $\log_{10}s_{\mathrm{1st}}\approx 4.856$, which agrees very well with the transition point shown in Fig. \ref{fig.E0123.a=0.2 and mu=12}.

\begin{figure}[h]
\begin{center}
\includegraphics[scale=0.55]{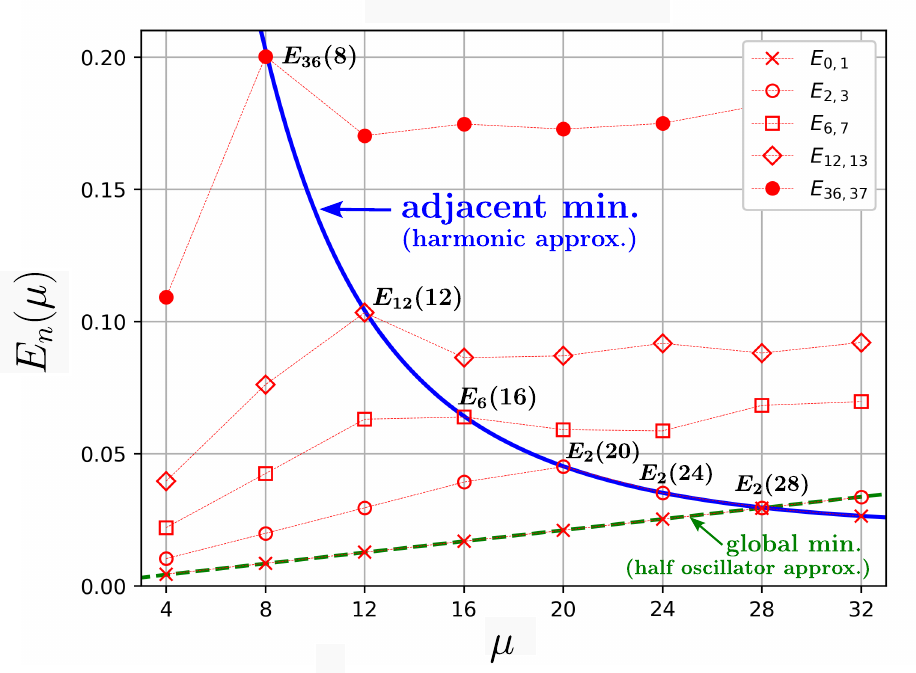}
\caption{Energy levels $E_n(\mu)$ of the $a=0.2$ system $(s=10^7)$. The discrete data points (red) are obtained via numerical diagonalization ($N_{\mathrm{dim}}=1000$) [connecting lines (dotted) are to guide the eye]. Dashed line (green) shows the zero-point energy of the global minima based on the half oscillator [see Eq. (\ref{eq.a=0.2.zpe of half oscillator})]. Solid curve (blue) shows our theoretical prediction of the energy of the adjacent minima [sum of Eqs. (\ref{eq.a=0.ZPE.harmonic approx}) and (\ref{eq. a=0.2. Vx1(mu)})]. The residual energies in Fig. \ref{fig.a=0.2.vary u} (right panel) are obtained using the data points indicated along this solid curve (as $E_{\mathrm{ref.}}$).}
\label{fig.En(mu)}
\end{center}
\end{figure}

Let us look at how the energy levels depend on $\mu$. Figure \ref{fig.En(mu)} shows the graphs of $E_n(\mu)$ when $s=10^{7}$, where the subscripts $n$ are indicated in the legend. Consider the ground state energy $E_{0,1}(\mu)$, plotted using crosses. The dashed line (green) shows the graph of $E_{\mathrm{wall}}(\mu)$ given by Eq. (\ref{eq.a=0.2.zpe of half oscillator}). We see that the numerical data agree closely with our theoretical prediction based on the half oscillator.

The energy levels corresponding to the adjacent minima were identified manually by inspecting the graphs of the energy eigenfunctions. These levels are indicated in Fig. \ref{fig.En(mu)} in boldface as `$E_n(\mu)$'. It is seen that the adjacent minima may not be the ground or first-excited states. For instance, when $\mu=16$, they are $E_{6,7}$, which is the third-excited state. The solid curve (blue) shows the zero-point energy of the adjacent minima obtained by adding Eqs. (\ref{eq.a=0.ZPE.harmonic approx}) and (\ref{eq. a=0.2. Vx1(mu)}) together. Once again, our theoretical description agrees very well with the numerical data.

To close this section, we briefly comment on our usage of the term `degenerate'. It is known as a theorem that in one dimension the ground state cannot be degenerate. So strictly speaking, our levels $E_{0,1}$ here exhibit an energy splitting which depends on the overlap between their wave functions. However, as can be seen from Fig. \ref{fig.E0123.a=0.2 and mu=12}, the two gaussian peaks are very far apart and their overlap is negligible. In our numerical diagonalization, this splitting of the ground state energy is not discernible in double precision. For convenience of discussion, we have therefore referred to the ground state as degenerate.

\subsection{Prevalence of diabatic transitions in quantum annealing}
\label{subsec.annealing.a=0.2}

\begin{figure}[h]
\begin{center}
\includegraphics[scale=0.55]{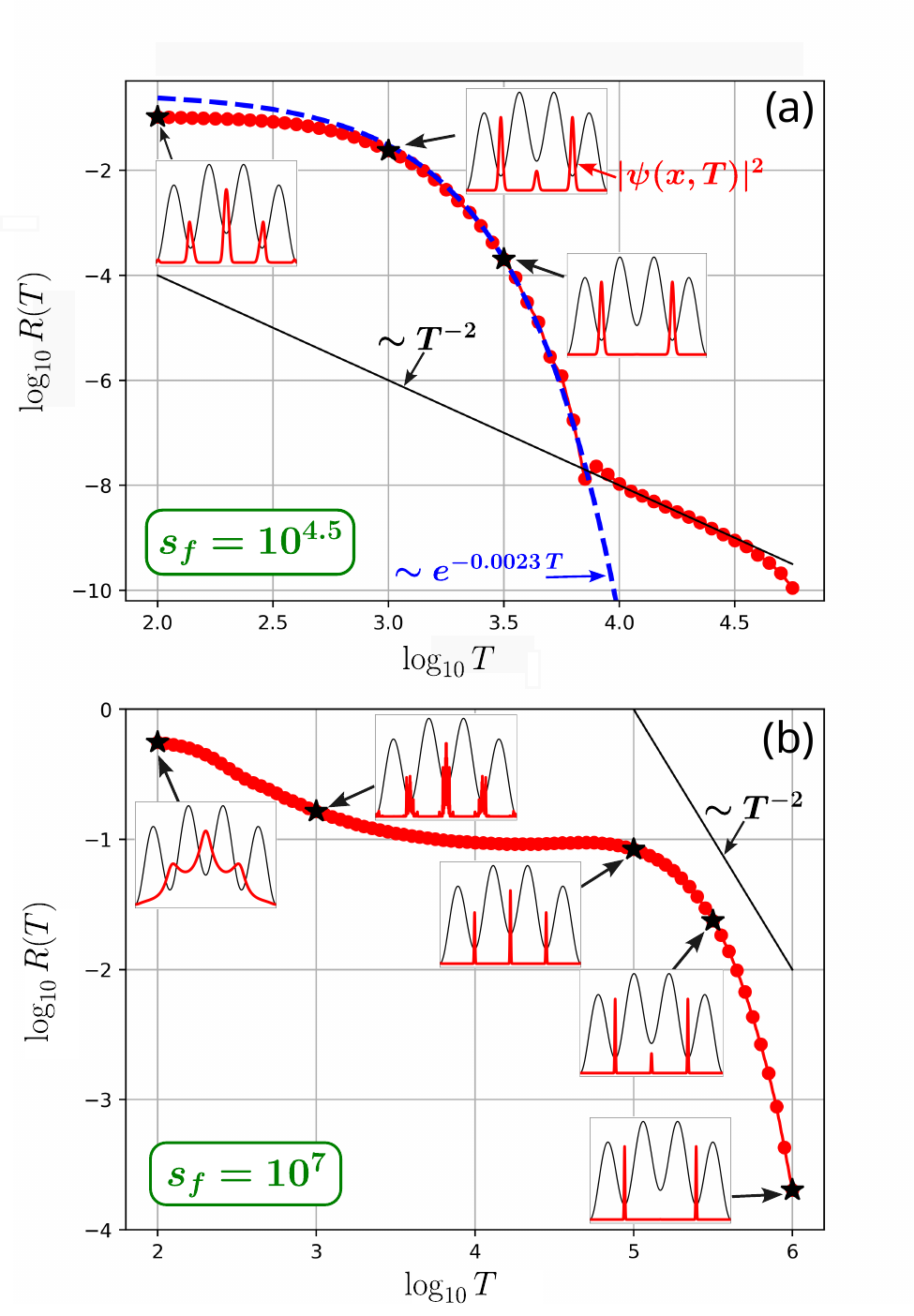}
\caption{Residual energy of the $a=0.2$, $\mu=8$ system under linear schedule ($s_i=1$). (a) Shallow annealing $s_f=10^{4.5}$. There is a crossover from exponential (dashed curve) to polynomial decay (straight line). Insets show the final probability densities $|\psi(x,T)|^2$ at some values of $T$ (stars). (b) Deep annealing $s_f=10^7$. One sees the emergence of an exponentially-slow `shoulder'. Simulation parameters are summarized in Table \ref{table. parameters. R(T)curves. a=0.2. linear}. Lines connecting data points are to guide the eye only.}
\label{fig.a=0.2.u=8.RE.two sfs}
\end{center}
\end{figure}

\begin{figure*}[t]
\begin{center}
\includegraphics[scale=0.55]{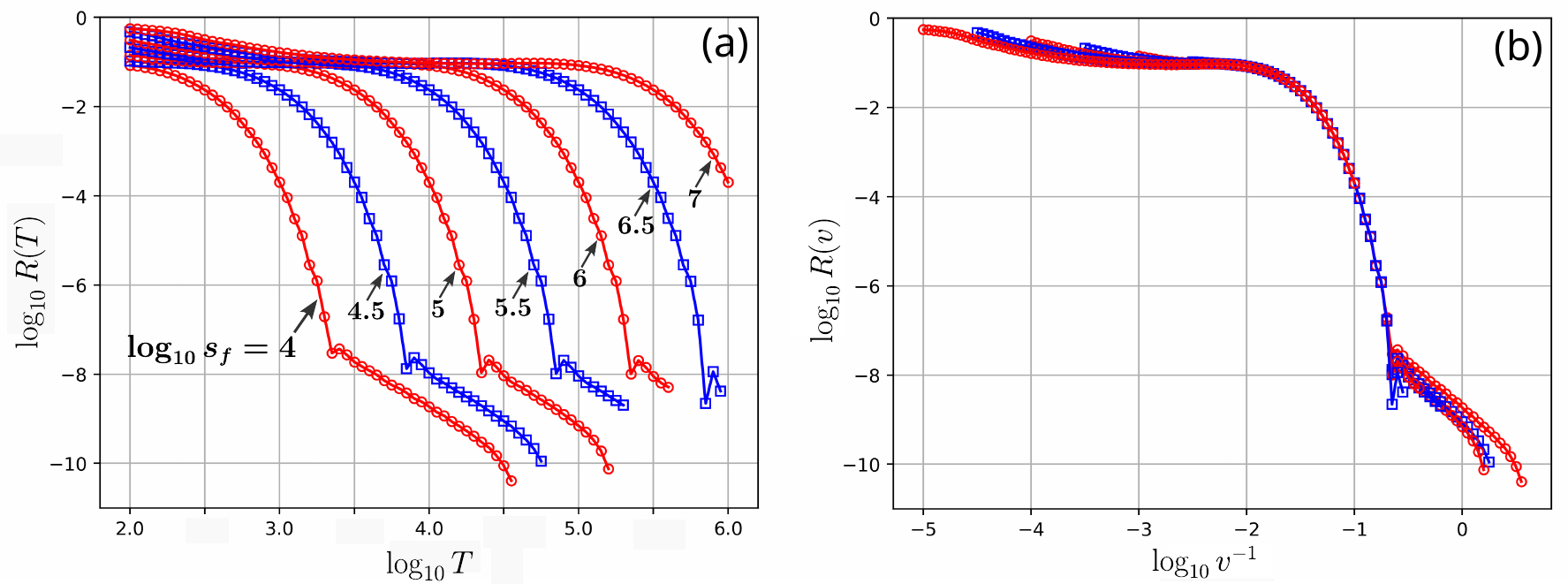}
\caption{(a) Dependence of $R(T)$ curves on annealing depth $s_f$. For $a=0.2$, $\mu=8$, under the linear schedule ($s_i=1$). As $s_f$ increases, the diabatic regime is prolonged, taking longer to reach the adiabatic regime (polynomial decay). (b) Replotting $R(T)$ in (a) as $R(v)$, where $v=\frac{s_f-s_i}{T}$ is the annealing speed [see Eq. (\ref{eq.linear schedule.definition})]. The curves for different $s_f$ now collapse onto one. Simulation parameters are summarized in Table \ref{table. parameters. R(T)curves. a=0.2. linear}. Lines connecting data points are to guide the eye only.}
\label{fig.a=0.2.u=8.vary sf}
\end{center}
\end{figure*}

\begin{figure*}[t]
\begin{center}
\includegraphics[scale=0.55]{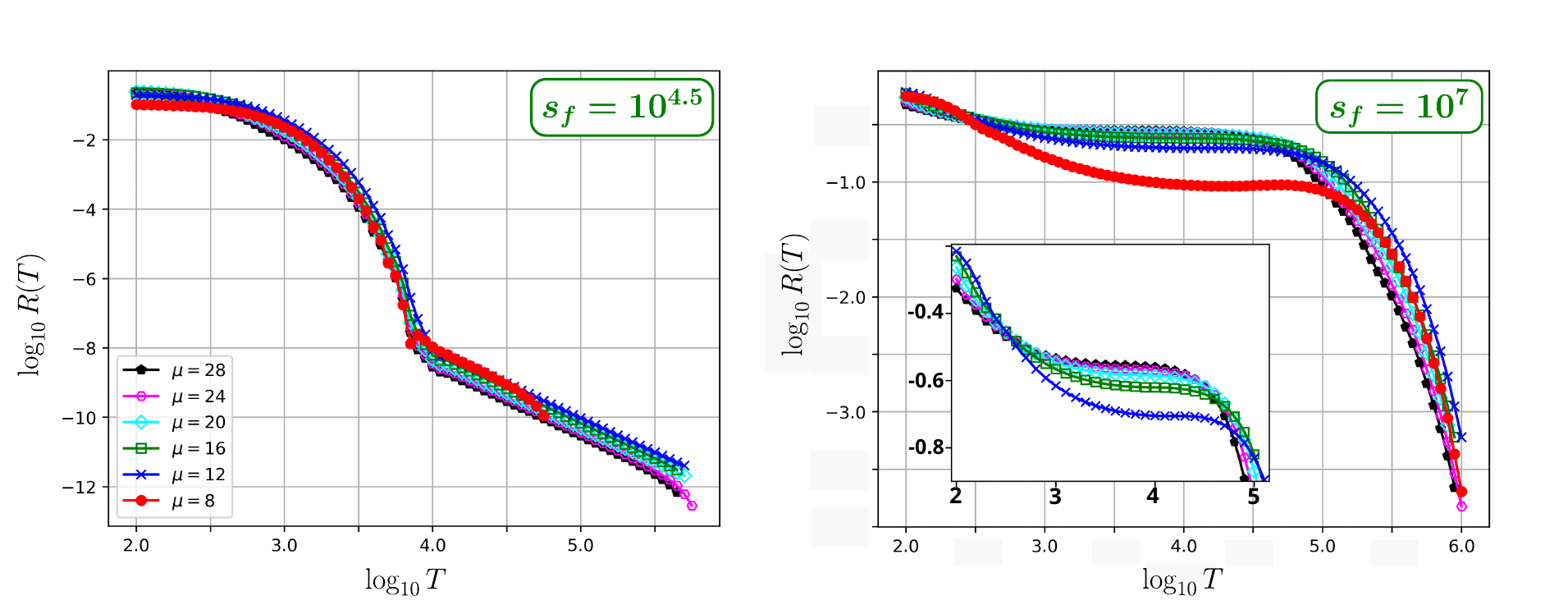}
\caption{Effects of increasing $\mu$ on the residual energy, for the $a=0.2$ system under linear schedule ($s_i=1$). Results from $\mu=8$ to 28 are shown. Symbols have same meanings in both panels. Left: Shallow annealing $(s_f=10^{4.5})$. Right: Deep annealing $(s_f=10^7)$. Inset: Close-up view of the early stages of the curves. In both panels, the curves for different $\mu$ are quite similar, exhibiting only slight dependence on $\mu$. Simulation parameters are summarized in Table \ref{table. parameters. R(T)curves. a=0.2. linear}. Lines connecting data points are to guide the eye only.}
\label{fig.a=0.2.vary u}
\end{center}
\end{figure*}

We now consider annealing. The simulation protocol is the same as in Sec. \ref{subsec.annealing.a=0}. Table \ref{table. parameters. R(T)curves. a=0.2. linear} summarizes the parameters used to obtain the residual energy curves in Figs. \ref{fig.a=0.2.u=8.RE.two sfs}, \ref{fig.a=0.2.u=8.vary sf}, and \ref{fig.a=0.2.vary u}. The horizontal lines separate information pertaining to different figures, for easy perusal. To utilize the table, first find the figure number under the first column, and then the specific parameters between the horizontal lines. We draw attention to the last column $E_{\mathrm{ref.}}$, noting that the target energy in Eq. (\ref{eq.Residual Energy.definition}) refers to the adjacent minima, not the wall minima.


\begin{table}[h]
\begin{center}
\begin{tabular}{ccccc}
\hline
\hline
\,\, Figure \, & \, $\mu$ \, &  \, $\log_{10}s_f$ \, & \, $N_{\mathrm{dim}}$ \, & \,\, $E_{\mathrm{ref.}}$ (at $s_f$) \,\,\, \\
\hline
\ref{fig.a=0.2.u=8.RE.two sfs}(a) & \multirow{2}{*}{8}  &  4.5 & 400   & $E_{2}(s_f)$   \\
\ref{fig.a=0.2.u=8.RE.two sfs}(b) &                     &  7   & 1000  & $E_{36}$  \\
\hline
\multirow{6}{*}{\ref{fig.a=0.2.u=8.vary sf}}  & \multirow{6}{*}{8}  &  4, 4.5 & 400                    & $E_{2}$ \\
                                              &                     &  5      & 600                    & $E_{4}$ \\
                                              &                     &  5.5    & \multirow{4}{*}{1000}  & $E_{6}$ \\
                                              &                     &  6      &                        & $E_{12}$ \\
                                              &                     &  6.5    &                        & $E_{20}$ \\
                                              &                     &  7      &                        & $E_{36}$ \\
\hline
\multirow{2}{*}{\ref{fig.a=0.2.vary u}, Left} & 8       & \multirow{2}{*}{4.5} & \multirow{2}{*}{400} & $E_{2}$ \\
                                              & $12-28$ &                      &                      & $E_{0}$ \\
\hline
\multirow{4}{*}{\ref{fig.a=0.2.vary u}, Right} & 8          & \multirow{4}{*}{7} & 1000 & $E_{36}$ \\
                                               & 12         &                    & 1000 & $E_{12}$ \\
                                               & 16         &                    & 1500 & $E_{6}$  \\
                                               & 20, 24, 28 &                    & 1500 & $E_{2}$  \\
\hline
\hline
\end{tabular}
\caption{Parameters used to obtain the $R(T)$ curves of the concave system $a=0.2$ under linear schedule shown in Figs. \ref{fig.a=0.2.u=8.RE.two sfs}, \ref{fig.a=0.2.u=8.vary sf}, and \ref{fig.a=0.2.vary u}. In all cases $s_i=1$.}
\label{table. parameters. R(T)curves. a=0.2. linear}
\end{center}
\end{table}


Figure \ref{fig.a=0.2.u=8.RE.two sfs} shows the $R(T)$ curves for the case of $\mu=8$. In both panels, the annealing parameter starts at $s_i=1$. In (a) the annealing ends at $s_f=10^{4.5}$, while in (b) it ends at $s_f=10^{7}$. For the rest of the paper, we shall refer to a small (large) $s_f$ as shallow (deep) annealing.

In panel (a), the residual energy shows a crossover from exponential to polynomial decay, as seen by comparing with the dashed curve (blue) and straight line (black). The $T^{-2}$ scaling indicates that one is in the adiabatic regime, discussed in Sec. \ref{subsec.annealing.a=0}. In this regime, the annealing is considered successful, as shown by the small residual energy [$R(T)<10^{-8}$]. On the other hand, an exponential decay is indication that the annealing suffers from diabatic transitions. The dashed curve $e^{-0.0023T}$ is fitted manually by hand. It is instructive to compare this with the Landau-Zener formula \cite{Landau32,Zener32}
\begin{equation}
P_{\mathrm{LZ}}=\exp 
\left( 
-\frac{2\pi \gamma^2}{\hbar v}
\right)
\label{eq.LZ.formula}
\end{equation}
where $P_{\mathrm{LZ}}$ is the diabatic transition probability, $2\gamma$ is the minimum energy gap at the avoided crossing, and $v$ is the asymptotic speed at which the two adiabatic curves approach and separate from each other. From Eq. (\ref{eq.linear schedule.definition}), we may approximate $v\propto T^{-1}$ for our purposes. We can relate Eq. (\ref{eq.LZ.formula}) to $R(T)$ via a similar reasoning used to derive Eq. (\ref{eq.app.R(T).step01}), which partly corroborates our finding that the residual energy decays exponentially with $T$. However, we were unable to obtain a quantitative agreement between the exponents of Eq. (\ref{eq.LZ.formula}) and $e^{-0.0023T}$. We believe that this is because some aspects of our model does not satisfy the assumptions underlying standard Landau-Zener theory. Specifically, the adiabatic curves here do not exhibit an avoided crossing, but rather, a `flat gap' in which a constant energy gap is maintained over a wide interval of $s$. This flat gap is easily seen if in Fig. \ref{fig.E0123.a=0.2 and mu=12} one plots $E_2(s)-E_0(s)$ along the vertical axis (in the region $\log s>2.5$). We shall discuss this in more detail when we examine the convex system in Sec. \ref{sec.a=-0.2}, where flat gap becomes the dominant feature. Although the exponential decay in Fig. \ref{fig.a=0.2.u=8.RE.two sfs}(a) is not described by the Landau-Zener paradigm, that diabatic transitions are indeed at play can nevertheless be evinced from the annealed wave functions. The insets in panel (a) show the final probability densities $|\psi(x,T)|^2$ for some values of $T$ (stars) along the $R(T)$ curve. When $T$ is small, the center local minimum is occupied. As $T$ increases, transition into the center minimum is gradually suppressed. Note that the center peak of $|\psi(x,T)|^2$ has already disappeared before the adiabatic regime is reached. This means that adiabaticity may not be necessary for QA to be effective.

Panel (b) shows the results of deep annealing ($s_f=10^7$). Compared to (a), here the residual energy decreases much slower (cf. their horizontal and vertical axes), and the exponential-polynomial crossover is not present. The system is clearly in the diabatic regime, as shown by the annealed wave functions in the insets. The shape of the $R(T)$ curve exhibits a `shoulder', which does not conform to the standard Landau-Zener formula Eq. (\ref{eq.LZ.formula}). To understand the relationship between the two curves of Fig. \ref{fig.a=0.2.u=8.RE.two sfs}, Fig. \ref{fig.a=0.2.u=8.vary sf}(a) shows the $R(T)$ curves for different values of $s_f$. We see that as $s_f$ increases, the diabatic regime is prolonged, taking longer to reach the adiabatic regime. This is because in Eq. (\ref{eq.linear schedule.definition}), increasing $s_f$ requires a corresponding increase in $T$ to maintain the same annealing speed. Let us define the annealing speed under the linear schedule as $v=\frac{s_f-s_i}{T}$. Figure \ref{fig.a=0.2.u=8.vary sf}(b) shows the $R(T)$ curves in panel (a) replotted as $R(v)$. The curves for different $s_f$ now collapse onto a single curve. The residual energy as a function of annealing speed $R(v)$ does not seem to depend on the annealing depth very much. Hence, in Fig. \ref{fig.a=0.2.u=8.RE.two sfs}(b) the diabatic-adiabatic crossover should appear if one continues the curve to longer $T$. Similarly, the shoulder in Fig. \ref{fig.a=0.2.u=8.RE.two sfs}(b) is caused by fast dynamics, and should also appear in Fig. \ref{fig.a=0.2.u=8.RE.two sfs}(a) if one extends the curve to shorter $T$. The actual physical mechanism underlying the shoulder is, however, unclear. Intra-basin excitations may be involved, because in the second inset ($T=10^3$) in Fig. \ref{fig.a=0.2.u=8.RE.two sfs}(b) the peaks of the final wave function exhibit short wavelength fluctuations (in the form of jaggedness). One way to analyze the problem theoretically may be through quenched dynamics \cite{Das06}.

Figure \ref{fig.a=0.2.u=8.vary sf}(a) also reveals some information about the computational complexity of QA. Let us denote $T_c$ as the point of crossover along each $R(T)$ curve. The data in Fig. \ref{fig.a=0.2.u=8.vary sf}(a) show that $T_c\propto s_f$; in other words, the computational cost of QA increases linearly with $s_f$. Note that this is more efficient than simulated annealing, whose annealing time $T_{\mathrm{SA}}$ increases exponentially with the final inverse temperature $\beta_f$ as $T_{\mathrm{SA}}\sim e^{c\beta_f}$ $(c>0)$ \cite{Nishimori.2001}. The parameter $s_f$ is also related to the precision of the $x$-coordinate of the minimized solution, measured by the width of the annealed wave function. For a harmonic oscillator with spring constant $k$ and mass $m$, the ground state density is a gaussian with standard deviation $\sigma_{D}=\sqrt{\frac{\hbar}{2\sqrt{km}}}\propto m^{-1/4}$. Substituting $m=s_f\propto T_c$, we obtain $\sigma_D \propto T_c^{-1/4}$. This is a rather expensive scaling for the precision of the annealed solution. On the other hand, the Boltzmann distribution of the harmonic oscillator is also a gaussian whose standard deviation scales with inverse temperature as $\sim \beta^{-1/2}$. This would suggest that in QA we define $s^2$ rather than $s$ as the annealing parameter, so as to have a similar scaling (of the precision) as in simulated annealing.

Hitherto in this section, we have been focusing on the case of $\mu=8$ where there are five minima on the potential surface. We now look at the effects of increasing the number of minima. Figure \ref{fig.a=0.2.vary u} shows the $R(T)$ curves as $\mu$ increases from 8 to 28. The left panel shows the results of shallow annealing, and the right one for deep annealing. For shallow annealing, the results for different $\mu$ are similar. There is no noticeable increase in crossover time $T_c$ as $\mu$ increases, meaning that increasing the number of minima does not lead to an increase in computational complexity. This is encouraging results for QA as a minimization algorithm, but also somewhat surprising since one would think that increasing the number of metastable traps on the energy landscape would increase the difficulty of optimization. There is also a visible displacement of the $R(T)$ curves vertically downwards as $\mu$ increases. In the adiabatic regime, this dependence on $\mu$ can once again be accounted for by the theory presented in Appendix \ref{app.R(T).a=0.adiabatic.derivation}. Following the same derivation leading to Eq. (\ref{eq.R(T).a=0.adaibatic approx}), one finds that $R_{\mathrm{adia.}}(T)$ now contains a multiplicative factor of $\mu^{-1}$,  due to the fact that the adjacent minima are two-fold rather than $(\frac{\mu}{2}-1)$-fold degenerate \cite{footnote.a=0.2.R(T).intrabasin excitations}. We verified that if we account for this factor, the data (in the adiabatic regime) in the left panel (Fig. \ref{fig.a=0.2.vary u}) indeed collapse onto a single straight line. 

In the right panel, the results for deep annealing are analogous, showing only a mild dependence on $\mu$. There is again no prolongation of the diabatic regime. The shoulder, observed for $\mu=8$, is also present. The later parts of the curves ($\log T>5$) are displaced slightly downwards with increase in $\mu$. On the shoulder, though, the trend is reversed. In the inset, one sees that as $\mu$ increases the curves are displaced upwards, approaching towards a limit for large $\mu$. 

To summarize this section, we studied how the residual energy of the concave system depends on $s_f$ and $\mu$, and found that these two parameters do not deteriorate the performance of QA in a serious way. We looked at $R(v)$ and saw how the curves of different annealing depths $s_f$ collapsed onto one. Since increasing $\mu$ (at fixed $s_f$) only results in mild changes to the $R(T)$ curves, we arrived at the above broad conclusion that the efficiency of QA is largely independent of these two factors. A second point concerns the nature of the $R(T)$ curve. We have seen that, in general, a $R(T)$ curve consists of adiabatic, diabatic, and fast (the shoulder) regimes. It might be worth noting that the diabatic regime is prevalent in all the curves we have presented here. In actual applications of QA, we feel that it would be quite difficult to attain the adiabatic regime, so annealings would generally suffer from diabatic transitions. In the diabatic regime, the residual energy lies mostly in the local minima, as seen from the insets in Fig. \ref{fig.a=0.2.u=8.RE.two sfs}. By contrast, in the adiabatic regime the residual energy is contained within the same minimum in the form of quanta excitations, as seen from the analysis in Appendix \ref{app.R(T).a=0.adiabatic.derivation}.

Lastly, we briefly explain why we left out the results for $\mu=4$ in our discussions. The minimum energy gap for $\mu=4$ is large enough to be overcomed for the annealing times we considered $(T\le 10^6)$. As its $R(T)$ curve is quite complicated and different from those of the other $\mu$'s, we omitted it to focus on the other curves' common and unifying features.


\section{Potential surface with convex envelope (\lowercase{\emph{a}}=$-$0.2)}
\label{sec.a=-0.2}

We now consider the third type of potential surface, the convex envelope obtained by setting $a=-0.2$ in Eq. (\ref{eq.Va(x).definition}). Figure \ref{fig.3 types V(x)}(c) shows the case when $\mu=24$. For this system, there is an unique global minimum at the center $x=\frac{L}{2}$. The shape of this surface is similar to that of the Rastrigin function, mentioned in the Introduction. It might be helpful to interpret this system vis-\`a-vis the previous concave one.  While the concave system looks at inter-basin annealing, this one focuses on intra-basin behavior, with the convex envelope playing the role of a large basin. Simulated annealing of Rastrigin-like models has been studied analytically by Shinomoto and Kabashima \cite{Shinomoto91}. Quantum annealing of similar models was subsequently investigated by Stella \emph{et al.} \cite{Stella05}, in which they adopted a coarsed-grained approximation to facilitate their analysis. These two pioneering works found that in simulated annealing the residual energy decays logarithmically as $\sim (\log T)^{-1}$, whereas in QA the decay follows a power law as $\sim T^{-\Omega}$. The exponent $\Omega$ is generally compatible with 1, but nevertheless dependent on specific details such as the particle's mass. In a more recent work, Koh and Nishimori \cite{Koh22} examined the QA of the Rastrigin function directly via numerical integration of the Schr\"odinger wave equation, without making any coarse-graining approximations. They verified certain aspects of Stella \emph{et al.}'s work (e.g., the power law), but also reported differences (e.g., with $\Omega \ge 2$ rather than 1). One limitation of Ref. \cite{Koh22} is that it did not elucidate how QA's performance depends on the ruggedness of energy landscape and annealing depth, which motivated our current work.

\subsection{Terrace of flat energy gaps and onset of wave function localizations}
\label{subsec.convex.a=-0.2.flat gaps}

\begin{figure}[t]
\begin{center}
\includegraphics[scale=0.55]{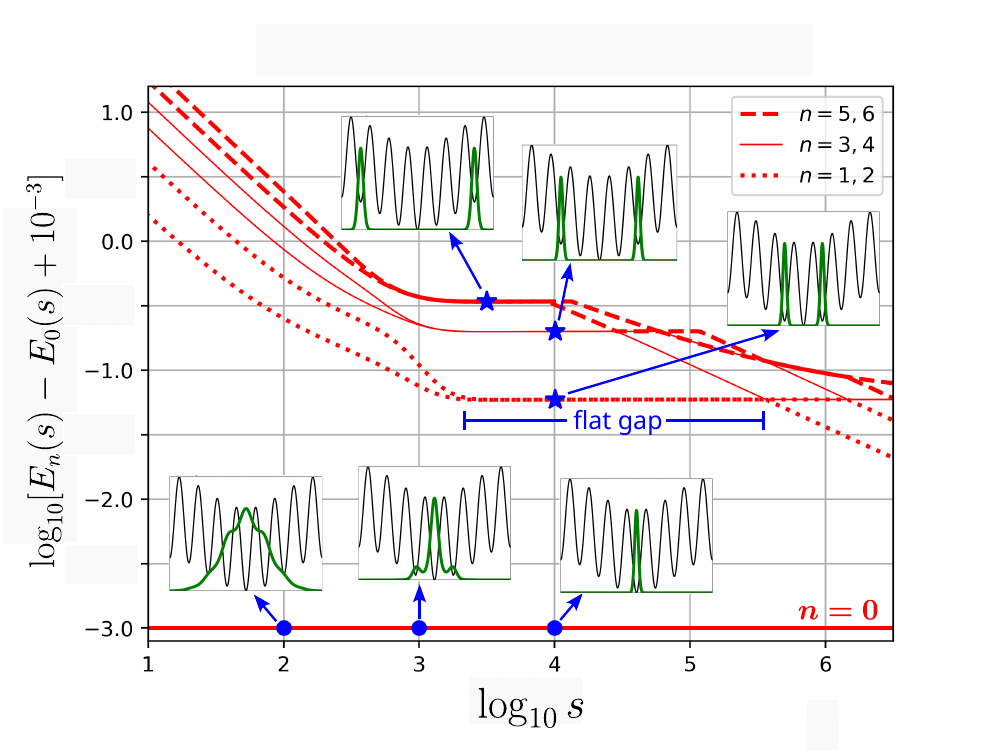}
\caption{Energy levels $E_{0-6}(s)$ of the $a=-0.2$, $\mu=16$ system ($N_{\mathrm{dim}}=1000$). The potential has a convex envelope and nine minima. Each level is shown relative to $E_0$ (plus a buffer $10^{-3}$), and represents the energy gap. The horizontal line at $-3$ represents $E_0$. The distinguishing feature is a series of plateaus, or `flat gaps', in which the gaps remain essentially constant over a wide range of $s$. The flat gap region between $E_{1,2}$ and $E_0$ is indicated in the figure. Insets show the probability densities of the energy eigenfunctions $|\psi_n(x)|^2$ at some representative values of $s$, indicated by the solid circles (for $n=0$) and stars (at flat gaps). The appearance of flat gaps indicates the onset of wave function localizations onto the minima of the potential surface. }
\label{fig.a=-0.2.flat gaps.u=16}
\end{center}
\end{figure}

We examine the energy levels first. Figure \ref{fig.a=-0.2.flat gaps.u=16} shows the seven lowest levels $E_{0-6}$ for the case of $\mu=16$, which has nine minima on the potential surface. To highlight the gap structure of this system, we plotted the energy difference $E_n(s)-E_0(s)$ along the vertical axis. (A buffer of $10^{-3}$ is being added so that $E_0$ can also be plotted.) The ground state is non-degenerate. As $s$ increases, due to the symmetry of the potential, $E_1$ and $E_2$ merge to become the doubly-degenerate first-excited state at $\log s\approx 3.3$ (dotted), while $E_{3,4}$ (thin line) and $E_{5,6}$ (dashed) merged slightly earlier at $\log s\approx 3.0$. Let us define the energy gap
\begin{equation}
\Delta_n=E_n-E_0
\label{eq.Delta n.definition}
\end{equation}
We see that after $E_{1,2}$ merged, the gap $\Delta_1$ remains essentially constant over the interval $\log s \in [3.3,5.5]$. The other gaps $\Delta_3$ and $\Delta_5$ also exhibit similar behaviors. To facilitate our discussion, let us refer to this constant energy gap as the `flat gap'. We briefly introduced the flat gap in Sec. \ref{subsec.annealing.a=0.2} when we discussed the concave system. Here in the convex system, flat gaps are prominent features in the `gap spectrum' $\Delta_n(s)$, forming a distinctive terrace-like structure with a series of plateaus.

The insets show the probability densities of the energy eigenfunctions $|\psi_n(x)|^2$ at some representative values of $s$, indicated by solid circles (for $n=0$) and stars (for $n=1,3,5$). For the ground state, one sees the familiar evolution from a broad delocalized form to a narrow gaussian localized at the global minimum. Note that in the second inset $(\log s=3)$, prior to the first-excited state becoming degenerate, the ground state density exhibits two small lobes at the two neigboring local minima. This is a harbinger that the energy eigenfunctions around the three lowest minima are about to be decoupled into separated, localized forms. We see this localization in the $n=1$ excited state (rightmost inset), whose probability density consists of two gaussians peaks located at the two neighboring minima. This localized form is attained as soon as $\psi_{1,2}(x)$ enter the flat gap, and is maintained thereafter. Similarly for $\psi_{3,4}(x)$ and $\psi_{5,6}(x)$, we see from the insets that their densities---along the flat gaps---are localized gaussians, centered at subsequently higher local minima.

\begin{figure}[h]
\begin{center}
\includegraphics[scale=0.55]{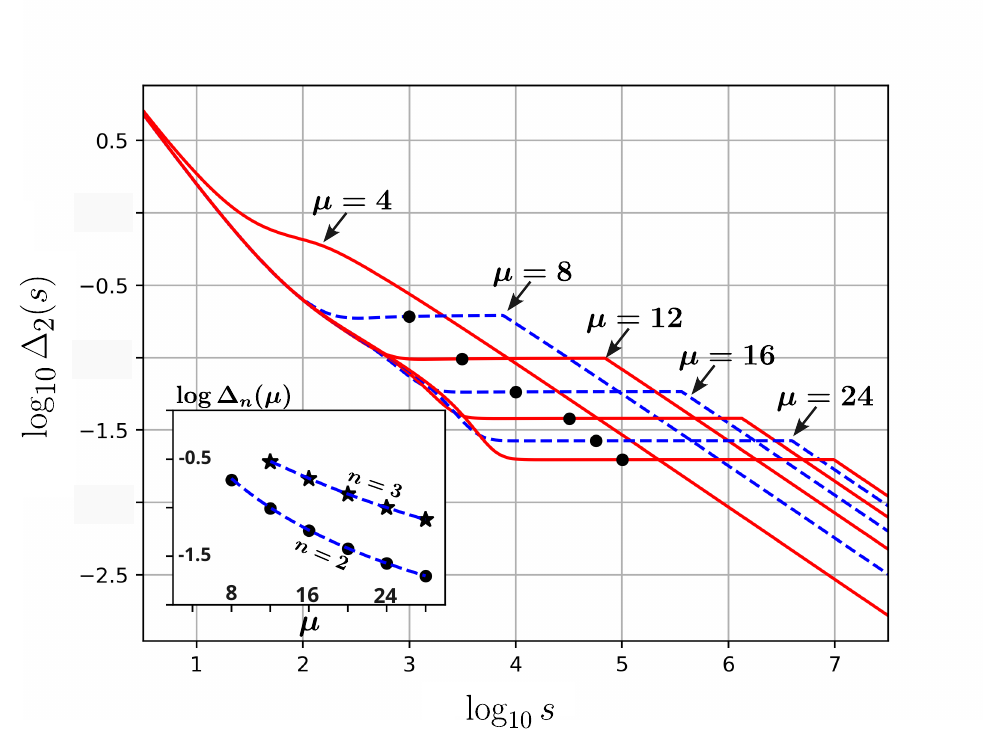}
\caption{Energy gap $\Delta_2(s)$ of the $a=-0.2$ system for $\mu$ from 4 to 28 ($N_{\mathrm{dim}}=1000$). As $\mu$ increases, the flat gap decreases and widens. The solid circles at the plateaus are replotted in the inset as $\Delta_2(\mu)$, and compared with our theoretical prediction Eq. (\ref{eq.Delta n wrt mu.theory}) (dashed curve). The stars show the corresponding comparison for $\Delta_3(\mu)$. }
\label{fig.convex.a=-0.2.Delta2.mu increases}
\end{center}
\end{figure}

In general, the overall features of the gap spectrum in Fig. \ref{fig.a=-0.2.flat gaps.u=16} are representative of other $\mu$ as well, albeit with some quantitative differences. Figure \ref{fig.convex.a=-0.2.Delta2.mu increases} shows the lowest gap $\Delta_2(s)$ for increasing values of $\mu$. One sees that the flat region becomes wider, lower, and shifts towards larger $s$ as $\mu$ increases. For the spectrum $\Delta_n(s)$ of each $\mu$, the number of plateaus also increases with increasing $\mu$ (i.e. more tiers) due to more minima on the potential surface.

Unlike the concave system, there is no energy gap closure in this system. Traditionally, a quantum phase transition is characterized by a vanishing energy gap between the ground and first-excited states \cite{Sachdev11}. As we are not dealing with a macroscopic system here, the notion of a phase transition may not be applicable. Nevertheless, one might still ask if it is possible to characterize the transition underwent by the particle. We would like to propose that the terrace of flat gaps in Fig. \ref{fig.a=-0.2.flat gaps.u=16} can serve as a sign of transition, analogous to a vanishing gap in the case of a quantum phase transition. The physical meaning of such a transition is a cascade of eigenfunction localizations into the many local minima on the potential surface, with the highest local minima localizing first and the global minimum last. Alternatively, one can also use the localizations of the lowest eigenstates $\psi_{0,1,2}(x)$ (e.g., at $\log s\approx 3.3$ in Fig. \ref{fig.a=-0.2.flat gaps.u=16}) to indicate (the end of) the cascade, although strictly speaking the transition unfolds over a range of $s$. We can interpret this as a quantum-to-classical transition, because the particle's wave functions become highly localized, which is what we normally associate with a classical particle. 

In simulated annealing, as the temperature traverses the critical point $\theta_c$, the free energy surface first broadens into a quartic minimum, and then bifurcates into a multi-valley landscape \cite{Fischer93}. If the system finds itself in one of the valleys at critical point, then it is very likely to continue staying in it thereafter. This is the general mechanism that gives rise to trappings by local minima in classical annealing. In the case of QA, we feel that the mechanism of trapping might be explained by the localization transition proposed above. In Fig. \ref{fig.a=-0.2.flat gaps.u=16}, notice that in the interval $3.3<\log s<4.0$, the lowest excited states $E_1$ to $E_6$ all have eigenfunctions which are localized at the local minima. This $s$-interval is analogous to $\theta_c$ in simulated annealing, with the terrace of flat gaps playing the role of the quartic Landau surface. When the system traverses this region of $s$ during QA, it is these localized states that will be excited with the highest probabilities (provided there is enough energy to do so). We feel that this gives an intuitive explanation of why in QA, diabatic transitions result in the wave function being trapped in local minima.

\subsection{Properties of the flat gap}
\label{subsec.a=-0.2.properties of L gap}

The occurrence of flat gaps is rather surprising, and we would like to understand whether is it a general phenomenon. The flatness of the lowest energy gap $\Delta_1$ was also reported in Ref. \cite{Koh22} for a wide range of parameters of the Rastrigin function Eq. (\ref{eq.Vsk(x).definition}). However, the underlying mechansim was not touched upon in that paper. In Appendix \ref{app.variational method to flat gap}, we provide an analysis from the perspective of variational method, to offer some insights into the mechanism. In the following, we discuss the main results from that analysis.

We saw in Fig. \ref{fig.a=-0.2.flat gaps.u=16} that along flat gaps the eigenfunctions are localized at the potential minima. Let us assume that they are gaussians of the form Eq. (\ref{eq.app.trial psi.gaussian}). Of relevance is the parameter $\alpha$ which controls the gaussian's width. The key result of Appendix \ref{app.variational method to flat gap} is that the gradient of the energy gap is given by  
\begin{equation}
\frac{\partial \Delta_n(s)}{\partial s}
\approx
-
\frac{\hbar^2}{4s^2}
\left[
\alpha_n(s)
-
\alpha_0(s)
\right]
\label{eq.gap gradient.wrt s}
\end{equation}
where the $n$th energy eigenstate $\psi_n(x)$ is a gaussian with $\alpha=\alpha_n$. The reason for the gap being `flat' is now clear. The gap gradient $\frac{\partial \Delta_n(s)}{\partial s}$ is, strictly speaking, non-zero, and depends on two factors. The first is the annealing parameter $s$ (i.e., the effective mass of the particle). As $s^2$ appears in the denominator, if $s$ is large, then the gap gradient will be very small. Indeed, in Fig. \ref{fig.a=-0.2.flat gaps.u=16} one sees that the flat gaps emerge when $s>10^3$. The second factor is the difference $\alpha_n(s)-\alpha_0(s)$. If the local curvature of the various minima on the potential surface are similar (as in the box model; see Fig. \ref{fig.3 types V(x)}), the localized gaussians will also have similar widths, so $\alpha_n\approx\alpha_0$, and this reduces the gap gradient further. On the other hand, if the curvatures differ from minimum to minimum, then this second factor would not play such an important role. For the box model (and the Rastrigin function), both factors contribute towards the gaps' flatness.

Equation (\ref{eq.gap gradient.wrt s}) originates from the kinetic energy operator and the localized gaussian ansatz Eq. (\ref{eq.app.trial psi.gaussian}). It does not depend on the potential energy of the system. Effects of boundary conditions should contribute only as minor perturbations to Eq. (\ref{eq.gap gradient.wrt s}). (We refer the reader to Appendix \ref{app.subsec.General validity} for discussions on the general validity of the result.) Hence, we believe that the appearance of flat gaps is a general phenomenon in the energy spectrum of continuous systems. It should occur so long as there are wave function localizations onto the (myriad) local minima of the potential surface.

For our box model, the magnitude of the flat gap can be obtained from potential differences. The zero-point energies are approximately the same at all the minima, and cancel when taking energy differences. From Fig. \ref{fig.a=-0.2.flat gaps.u=16}, one sees that the $m$th flat gap $\tilde{\Delta}_m$ is associated with the potential minimum located at $x_m=\frac{L}{2}\pm\frac{2mL}{\mu}$. Taking the difference $V_{\mathrm{box}}(x_m)-V_{\mathrm{box}}(x_0)$, we obtain 
\begin{equation}
\tilde{\Delta}_m(\mu)
=
a
\left[
\cos\left(
\frac{4m\pi}{\mu}
\right)
-1
\right]
\label{eq.Delta n wrt mu.theory}
\end{equation}
Figure \ref{fig.convex.a=-0.2.Delta2.mu increases} compares Eq. (\ref{eq.Delta n wrt mu.theory}) with exact numerical results. The values of $\Delta_2$ at the plateaus (circles) are replotted in the inset as $\Delta_2(\mu)$. The case for $\Delta_3(\mu)$ is also shown (stars). The dashed curves in the inset show the graphs of $\tilde{\Delta}_1(\mu)$ and $\tilde{\Delta}_2(\mu)$. It is seen that the classical result Eq. (\ref{eq.Delta n wrt mu.theory}) describes the flat gap very well.

\subsection{Annealing under linear schedule}
\label{subsec.a=-0.2.annealing results}

\begin{figure}[h]
\begin{center}
\includegraphics[scale=0.55]{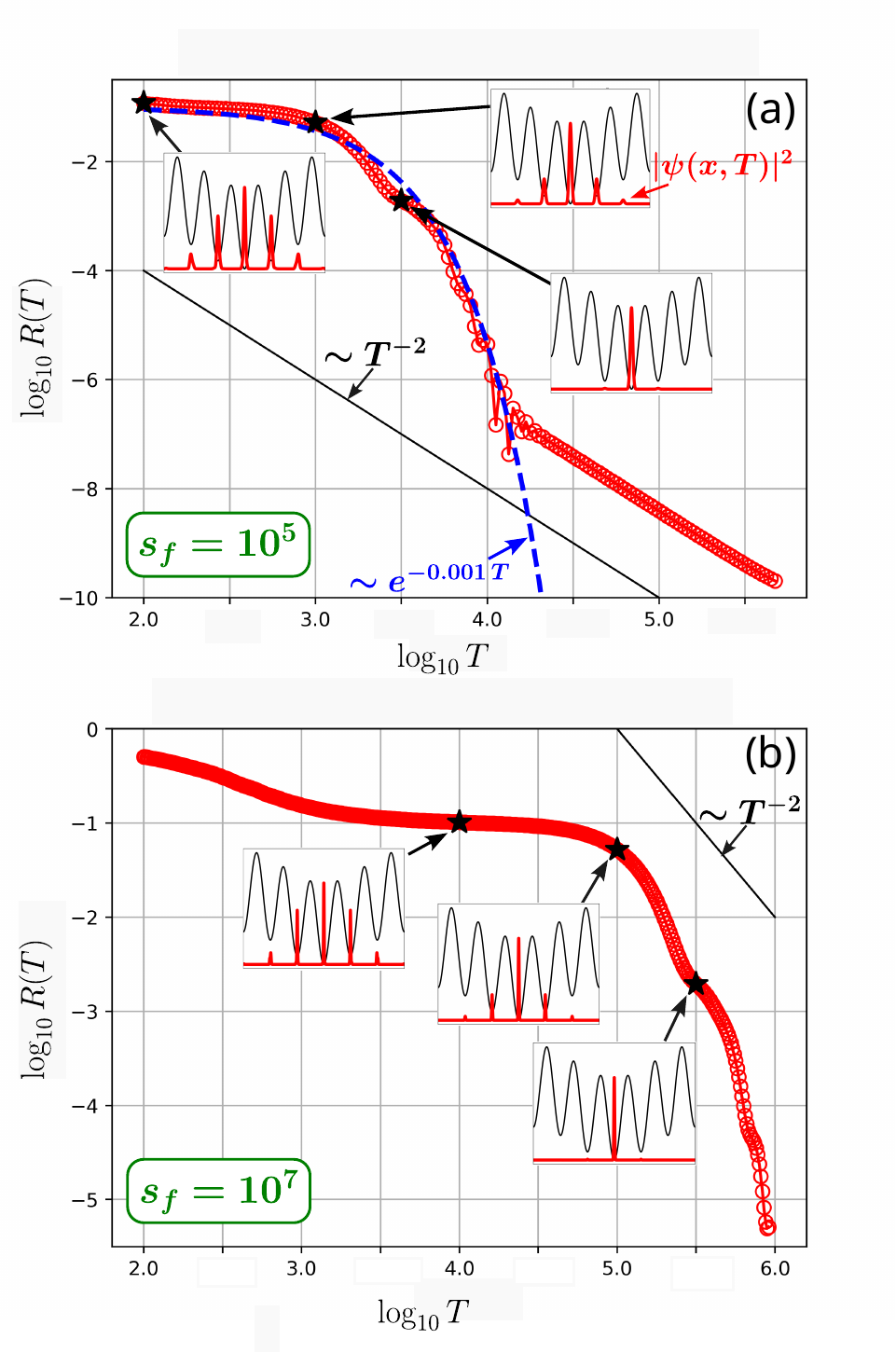}
\caption{Residual energy of the convex system $a=-0.2$, $\mu=12$ under linear schedule [$s_i=10^{0.5}$, $N_{\mathrm{dim}}=500$, $E_{\mathrm{ref.}}=E_0(s_f)$]. (a) Shallow annealing $s_f=10^5$. (b) Deep annealing $s_f=10^7$. Insets show the final probability densities $|\psi(x,T)|^2$ at the indicated values of $T$ (stars). Overall, the results are similar to those of the concave system in Fig. \ref{fig.a=0.2.u=8.RE.two sfs}.}
\label{fig.a=-0.2.convex.R(T).shallow and deep.}
\end{center}
\end{figure}

\begin{figure*}[t]
\begin{center}
\includegraphics[scale=0.55]{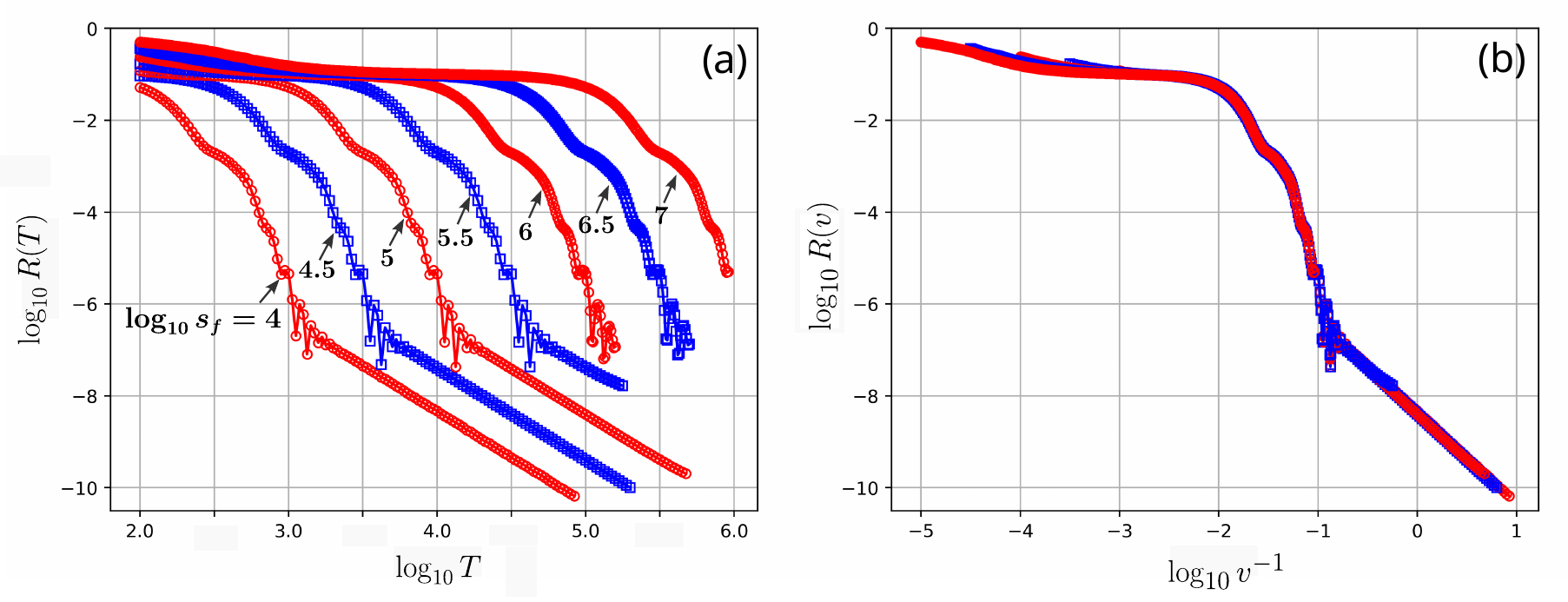}
\caption{(a) Dependence of $R(T)$ on annealing depth $s_f$. For the $a=-0.2$, $\mu=12$ system under linear schedule [$s_i=10^{0.5}$, $N_{\mathrm{dim}}=500$, $E_{\mathrm{ref.}}=E_0(s_f)$]. (b) Replotting $R(T)$ as $R(v)$, as in Fig. \ref{fig.a=0.2.u=8.vary sf}(b). Overall, the behavior is similar to that of the concave system shown in Fig. \ref{fig.a=0.2.u=8.vary sf}.}
\label{fig.a=-0.2.vary sf}
\end{center}
\end{figure*}

\begin{figure*}[t]
\begin{center}
\includegraphics[scale=0.55]{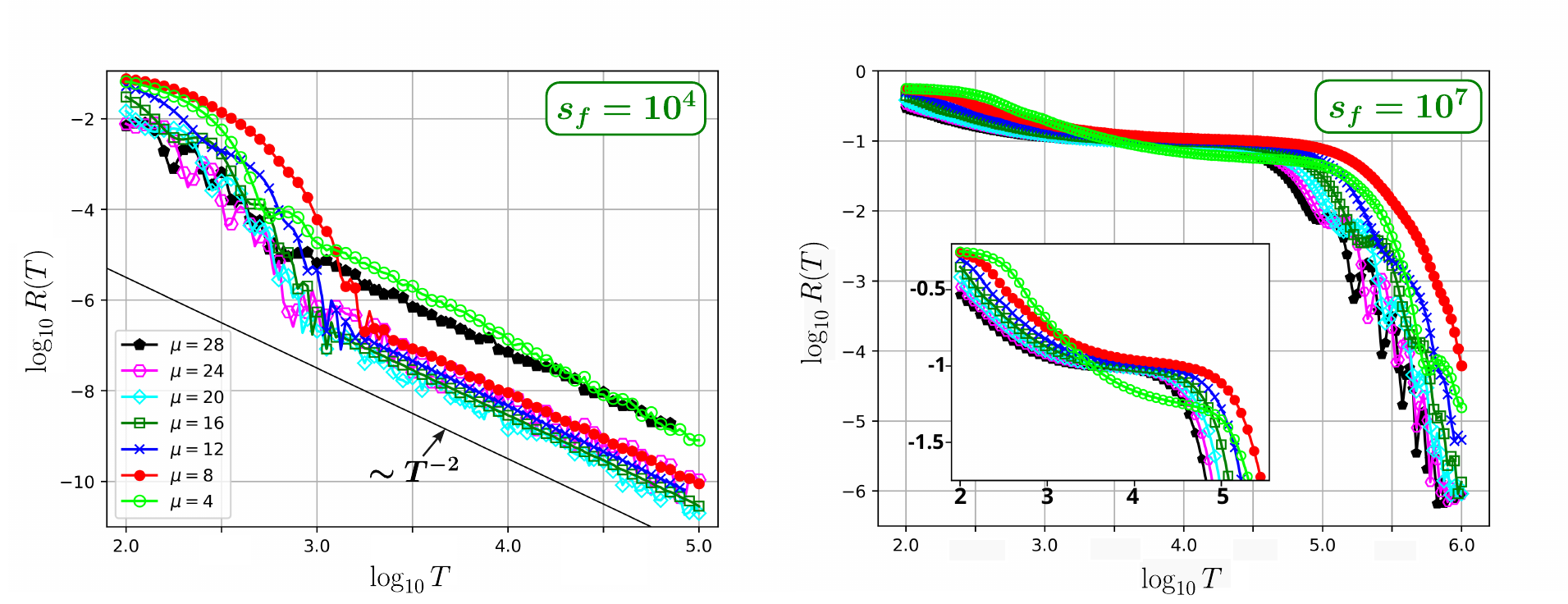}
\caption{Effects of increasing $\mu$ on the residual energy, for the $a=-0.2$ system under linear schedule [$s_i=10^{0.5}$, $E_{\mathrm{ref.}}=E_0(s_f)$]. The figure is organized similar to Fig. \ref{fig.a=0.2.vary u}. Left: Shallow annealing $s_f=10^{4}$ ($N_{\mathrm{dim}}=500$). Right: Deep annealing $s_f=10^7$ ($N_{\mathrm{dim}}=1000$). The behavior is similar to the concave system, in that increasing the number of local minima does not affect $R(T)$ seriously.}
\label{fig.a=-0.2.vary u}
\end{center}
\end{figure*}

We now consider annealing. The simulation protocol is again the same as in Sec. \ref{subsec.annealing.a=0}. For this system, the reference energy $E_{\mathrm{ref.}}$ in Eq. (\ref{eq.Residual Energy.definition}) refers to the ground state located at the global minimum $x=\frac{L}{2}$. Simulation parameters are given in the captions of Figs. \ref{fig.a=-0.2.convex.R(T).shallow and deep.}, \ref{fig.a=-0.2.vary sf}, and \ref{fig.a=-0.2.vary u}.

Overall, the results of the convex system are similar to those of the concave one. Figure \ref{fig.a=-0.2.convex.R(T).shallow and deep.} shows the $R(T)$ curves of $\mu=12$ for $s_f=10^5$ and $10^7$. In panels (a) and (b), one again sees the crossover and the shoulder (as in Fig. \ref{fig.a=0.2.u=8.RE.two sfs}). The insets show the final probability densities $|\psi(x,T)|^2$ at some values of $T$ (stars). We see that the wave function is trapped in local minima if annealing time is short. As mentioned in Sec. \ref{subsec.convex.a=-0.2.flat gaps}, we think that these diabatic excitations can be explained by the system traversing the flat gap region.

The dashed curve $e^{-0.001T}$ in panel (a) is fitted manually by hand to the diabatic portion of the $R(T)$ curve. As in Fig. \ref{fig.a=0.2.u=8.RE.two sfs}(a), here we are also unable to obtain a reasonable agreement between the exponents of the Landau-Zener formula Eq. (\ref{eq.LZ.formula}) and $e^{-0.001T}$. As mentioned before, in Landau-Zener theory the energy gap is assumed to be an avoided crossing, shaped like a `V'. Here, however, our system exhibits a flat gap, shaped like an `L'. We think that this is the main reason why Eq. (\ref{eq.LZ.formula}) does not account for our simulation data.

Figure \ref{fig.a=-0.2.vary sf} shows the effects of annealing depth $s_f$ on the residual energy. The figure is organized similar to Fig. \ref{fig.a=0.2.u=8.vary sf} of the concave system, and the results are similar as well. The $R(T)$ curves of various $s_f$ collapse onto one when replotted as a function of annealing speed $R(v)$. The comments concerning computational complexity and scalings of QA in Sec. \ref{subsec.annealing.a=0.2} are therefore applicable to the current convex surface as well.

As our last item, we look at the dependence on the number of local minima. Figure \ref{fig.a=-0.2.vary u} shows the $R(T)$ curves as $\mu$ increases from 4 to 28. In this case, there are similarities but also differences from the concave system (see Fig. \ref{fig.a=0.2.vary u}). In both (a) and (b), one sees that there is no significant prolongation of the diabatic regime with increase in $\mu$. This means that the performance of QA is not seriously affected by the number of local minima, which is again encouraging news for QA as an optimization algorithm. We see some vertical displacement of the curves with $\mu$. For the $T^{-2}$ regime in the left panel, the adiabatic theory presented in Appendix \ref{app.R(T).a=0.adiabatic.derivation} again offers some insights. Repeating the same derivation, noting that this time the ground state (at $x=\frac{L}{2}$) is non-degenerate, one finds that $R_{\mathrm{adia.}}(T)$ contains a multiplicative factor of $\mu^{-1}$, like the concave system. This accounts for the downward displacement of the curves until $\mu=20$. 

The curves of $\mu=24$ and 28 (left panel) are displaced upwards, contrary to the downward trend. We examined their ground state eigenfunctions $\psi_0(x)$ at $s_f=10^4$, and found that the probabilities are significant at the two neighboring minima. In other words, the particle is delocalized among the three lowest minima. This means that in Eq. (\ref{app.eq.<2|Hdot|0>}) the connection of the operator $p^2$ between the global and local minima must be included when calculating the matrix element. We think that this is why $\mu=24$ and 28 are slightly different from the rest. As a casual observation, note that the need to include such `long-ranged' connections blurs the line between intra-basin (adiabatic) and inter-basin (diabatic) transitions, which is perhaps the reason why the diabatic-adiabatic crossover is not very distinct for these two $\mu$'s.


\section{Conclusion and discussions}
\label{sec.conclusion}

In this paper, we proposed a box model of continuous space quantum annealing. This model facilitates the numerical simulation of QA in that one can change from energy to coordinate representation without involving special functions (which may suffer from numerical overflows at high quantum numbers). Our setup enables us to study two aspects of continuous space QA which have previously not been examined in detail, namely how the number of local minima on the potential surface (i.e. landscape ruggedness) and the annealing depth (i.e. precision of solution) affect the performance of QA. Both require the use of large basis sets during simulations. We investigated three types of potential energy landscapes, those with a flat, a concave, or a convex envelope. For each type of landscape, we first studied the static behavior by examining the low-lying energy spectrum, followed by dynamics using residual energy as a measure of annealing performance.

For statics, the concave system exhibits an energy gap closure, resulting in a first-order transition of the ground state. For the convex system, the main feature is a terrace of flat gaps, discussed at length in Secs. \ref{subsec.convex.a=-0.2.flat gaps} and \ref{subsec.a=-0.2.properties of L gap}. They are also present in the concave system, and we believe them to be a ubiquitous feature in the gap spectrum of potentials with multiple local minima. Their physical interpretation is a cascade of eigenfunction localizations onto the local minima of the potential landscape. We also proposed that the flat gap region may be responsible for the wave function being trapped in local minima during QA. Lastly, in all three types of systems, we showed how the energy levels and gaps can be understood quite intuitively in terms of just zero-point energy and the classical potential.

In dynamics, our main result is that the performance of QA is not affected very much by the ruggedness of the landscape and the depth of annealing. This is true for all three types of landscapes. Generally speaking, a $R(T)$ curve exhibits a three-regime behavior: a fast regime (shoulder), a diabatic regime (exponential decay), and an adiabatic regime ($T^{-2}$ decay). We showed how the $R(T)$ curves of different $\mu$ and $s_f$ collapse onto one when replotted as a function of annealing speed $R(v)$. Overall, we think that the results in this work indicate that QA is very promising as an algorithm for optimization in continuous space.

It might be helpful to relate our work to optimization problems in general. In realistic systems (e.g. cluster and protein molecules \cite{Miller99,Wales18}), the energy landscapes are more complex than the ones studied here. We can think of our three types of landscapes as motifs. By studying these smaller and more manageable cases, we can build towards understanding general and complex ones, many of which are numerically intractable. One thing we can do is modify a motif slightly to increase the difficulty of the problem. For instance, in the concave system we can lean the envelope to one side, thereby lifting the degeneracy of the ground state. The particle will then have to overcome small as well as large (furnished by the envelope) energy barriers to attain the the global minimum. Another possibility is to combine two or more motifs into one. All these can be readily implemented using our box model, since any potential function can be expanded as a Fourier series.

As reviewed in the Introduction, Stella \emph{et al.} reported that for the Shinomoto-Kabashima potential the residual energy decays as $T^{-\Omega}$ where $\Omega=\frac{1}{3}$ (but is in general dependent on circumstances, and compatible with 1) \cite{Stella05}. Inack and Pilati found the same decay exponent \cite{Inack15}. Koh and Nishimori reported that $\Omega=2$ (under most circumstances, using the linear schedule), but also found situations whereby the decay is exponential in the manner reported here \cite{Koh22}. Perhaps one contribution of this work is to foreground the ubiquity of the exponential-polynomial crossover, and accentuate the prevalence of diabatic transitions in finite time QA. This does not seemed to have received much emphasis in the literature, at least in the context of continuous space QA. As we have seen in many of the $R(T)$ curves being presented, one usually do not have the luxury to anneal until the adiabatic regime. Although the Landau-Zener theory explains the diabatic regime of QA in a two-level system very well \cite{Morita07}, here we have seen that it is unable to provide a quantitative description of our simulation data. On a separate but related note, the point of diabatic-adiabatic crossover $T_c$ bears some resemblence to a second-order phase transition in statistical mechanics. It would be interesting to inquire how it can be explained theoretically.

\textcolor{black}{
From the evolving of the wave function and of the residual energy 
during the linear annealing schedule, 
we could learn that, given the total annealing duration $T$ and the final parameter  $s_f$,
an optimized annealing protocol in practical application 
should probably contain three stages: an initial stage of rapid growth of $s$, 
a very slow stage in the middle, and an adiabatic final stage. 
The whole process $s(t)$ might roughly resemble the curve obtained by rotating the curve in Fig.~\ref{fig.a=-0.2.convex.R(T).shallow and deep.}(a) by 90 degrees counterclockwise.
The initial stage is mainly to make that the overall shape of the wave function 
can reflect the ground state of the time-dependent Hamiltonian, 
the middle stage represents the diabatic process to release the residual energy, 
and the final stage is for adiabatic annealing.
}

Lastly, let us comment on some directions for future studies. It would be interesting to generalize the model to two dimensions. One possibility is to consider a radially symmetric potential (e.g., rotate Eq. (\ref{eq.Vbox.definition}) about the $z$-axis). The advantage of such a setup is that the equations of motion are separable in radial and angular coordinates, so the computation is still effectively one-dimensional. Another way is to consider a box version of the two-dimensional Rastrigin function. In this case, the required number of basis functions will scale quadratically relative to our current system, greatly increasing the cost of computation. In return, the energy landscape is also much richer than a radially symmetric one. Quantum annealing in two dimensions is also interesting for another reason: tunneling. The role played by tunneling in QA has been studied for the one-dimensional Rastrigin function \cite{Koh22}. One-dimensional tunneling is relatively simple in the sense that there is only one direction for the particle to maneuver. By contrast, tunneling in two dimensions is highly non-trivial because there are many directions in which tunneling can proceed, and this may have important effects on QA's performance (e.g. a different rate of convergence). It would also be interesting to visualize how a particle tunnels towards the global minimum in a two-dimensional milieu. We hope to address these and related issues in our future works.

\begin{acknowledgements}
We acknowledge the support by the National Natural Science Foundation of China (NSFC) under Grant No.~12275263, as well as the Innovation Program for Quantum Science and Technology (under Grant No. 2021ZD0301900). YD is also supported by the Natural Science Foundation of Fujian Province 802 of China (Grant No. 2023J02032).
\end{acknowledgements}


\appendix


\section{Derivation of residual energy under adiabatic approximation}
\label{app.R(T).a=0.adiabatic.derivation}

In this appendix, we derive Eq. (\ref{eq.R(T).a=0.adaibatic approx}), the residual energy $R(T)$ of the $a=0$ system under adiabatic approximation. We work under the assumption that the annealing is sufficiently slow, so only the excited state closest to the ground state is activated. This means that the wave function $\psi$ at the end of annealing (i.e., at $t=T$) takes the form
\begin{equation}
\psi(T)
\approx
\sum_{\lambda}
c_0^{\lambda}(T) \, |0^{\lambda}\rangle
+
\sum_{\lambda}
c_2^{\lambda}(T) \, |2^{\lambda}\rangle
\label{eq.app.psi(T).assumption}
\end{equation}
where the superscript $\lambda$ ($=1,\cdots,\frac{\mu}{2}-1$) labels all the minima of $V_{\mathrm{box}}(x)$ except the two at the walls. The kets $|0^{\lambda}\rangle$ and $|2^{\lambda}\rangle$ refer to the ground and second excited states of the $\lambda$th minimum at $s_f$, which we take to be harmonic oscillator energy states. We chose $|2^{\lambda}\rangle$ and not $|1^{\lambda}\rangle$ because, as we shall see below, the kinetic energy operator $p^2$ does not connect $|1^{\lambda}\rangle$ to $|0^{\lambda}\rangle$. Inserting Eq. (\ref{eq.app.psi(T).assumption}) into Eq. (\ref{eq.Residual Energy.definition}), utilizing the normalization of $\psi(T)$, and defining 
\begin{equation}
\Delta_{20} = E_2(s_f)-E_0(s_f)
\label{}
\end{equation}
one has
\begin{equation}
R(T)
\approx
\Delta_{20}
\,
\sum_{\lambda}
|c_2^{\lambda}(T)|^2
=
\Delta_{20}
\,
\left(
\frac{\mu}{2}-1
\right)
\,
|c_2(T)|^2
\label{eq.app.R(T).step01}
\end{equation}
In the second step of Eq. (\ref{eq.app.R(T).step01}), the label $\lambda$ is dropped since all the minima are identical. In the following, we shall drop $\lambda$ from $|0^{\lambda}\rangle$ and $|2^{\lambda}\rangle$ as well.

The starting point of our derivation is the probability amplitude $c_2(T)$, which is given by adiabatic approximation as (see, for instance, Ref. \cite{Ballentine98})
\begin{equation}
c_2(T)
\approx
-i\hbar
\,
\frac{
\langle 2|\dot{H}|0\rangle
}
{(\Delta_{20})^2}
\,
\left[
\exp\left(\frac{iT\Delta_{20}}{\hbar}\right)
-
1
\right]
\label{eq.app.c2(T).adiabatic formula}
\end{equation}
where $\dot{H}=\frac{dH[s(t)]}{dt}$. For the linear schedule Eq. (\ref{eq.linear schedule.definition}), one has
\begin{equation}
\langle 2|\dot{H}|0\rangle
=
-
\frac{1}{2mT}
\left(
\frac{s_f-s_i}{s_f^2}
\right)
\langle 2|p^2|0\rangle
\label{app.eq.<2|Hdot|0>}
\end{equation}
To evaluate $\langle 2|p^2|0\rangle$, we write
\begin{equation}
p=
i\sqrt{\frac{m_f\hbar\omega_f}{2}}
(a^{\dagger}-a)
\label{eq.app.p=a and a+}
\end{equation}
where $m_f=ms_f$ and 
\begin{equation}
\omega_f=\frac{\pi\mu}{\sqrt{2ms_f}L}
\label{eq.app.omegaf.definition}
\end{equation}
are, respectively, the effective mass and angular frequency of one of the minima at $s_f$. The creation and annihilation operators in Eq. (\ref{eq.app.p=a and a+}) can thus act on $|0\rangle$ and $|2\rangle$. Finally, we need
\begin{equation}
\Delta_{20} = 2\hbar\omega_f
\label{eq.app.D20.SHO approx}
\end{equation}
Inserting Eqs. (\ref{eq.app.c2(T).adiabatic formula}) to (\ref{eq.app.D20.SHO approx}) into Eq. (\ref{eq.app.R(T).step01}), we obtain Eq. (\ref{eq.R(T).a=0.adaibatic approx}).


\section{Variational method analysis of flat energy gap}
\label{app.variational method to flat gap}

In this appendix, we analyze the flat energy gap using the variational method. In Ref. \cite{Koh22}, numerical results showing the existence of flat gaps in the Rastrigin system [Eq. (\ref{eq.Vsk(x).definition})] were presented. However, the underlying mechanism was not touched upon in that paper. Here, we analyze the phenomenon from the perspective of variational method, to offer some insights into its mechanism. Although it is possible to study the box model, we prefer to revisit the Rastrigin function instead because the gaussian integrals can be performed analytically, making the derivations simpler and the analysis clearer. We shall see that the main features of the flat gap does not depend on boundary conditions. To make our discussion succinct, the reader is referred to Ref. \cite{Koh22} for implicit information concerning the Rastrigin system (its graph, parameters etc).


\subsection{Variational energy of Rastrigin system}
\label{}

In the variational method \cite{Ballentine98}, one proposes a trial wave function $\psi^{\mathrm{var}}(x;\vec{\alpha})$ where $\vec{\alpha}$ are variational parameters chosen to minimize the expectation energy, i.e.
\begin{equation}
\tilde{E}^{\mathrm{var}}
=
\min_{\vec{\alpha}}
\,
\left\langle
\psi^{\mathrm{var}}(\vec{\alpha})|H|\psi^{\mathrm{var}}
(\vec{\alpha})
\right\rangle
\label{eq.app.variational method.definition}
\end{equation}
where $\tilde{E}^{\mathrm{var}}$ is the approximation to the targeted energy. For our purposes, the Hamiltonian $H$ is given by Eq. (\ref{eq.H(s).definition}), with $V_{\mathrm{box}}(x)$ replaced by the Rastrigin function Eq. (\ref{eq.Vsk(x).definition}). For the wave function, we adopt the following gaussian ansatz
\begin{equation}
\psi^{\mathrm{var}}(x;\alpha,x_0)=
\left(\frac{\alpha}{\pi}\right)^{\frac{1}{4}}
\exp
\left[
-\frac{\alpha}{2}
(x-x_0)^2
\right]
\label{eq.app.trial psi.gaussian}
\end{equation}
where $x\in(-\infty,\infty)$, and the variational parameters $\alpha$ and $x_0$ control the width and center of the gaussian, respectively. Equation (\ref{eq.app.trial psi.gaussian}) is normalized. Let us denote
\begin{equation}
E^{\mathrm{var}}(\alpha,x_0)
=
\left\langle
\psi^{\mathrm{var}}|H|\psi^{\mathrm{var}}
\right\rangle
\label{app.Evar.denotation}
\end{equation}
Substituting Eq. (\ref{eq.app.trial psi.gaussian}) into Eq. (\ref{app.Evar.denotation}), one obtains
\begin{eqnarray}
E^{\mathrm{var}}(\alpha,x_0)
&\!&=
\frac{\hbar^2}{2m}
\left(
\frac{\alpha}{2}
\right)
+
\frac{k}{2}
\left[
\frac{1}{2\alpha}
+
x_0^2
\right]+
\nonumber \\ & &
\frac{h_0}{2}
\left[
1
-
\cos
\left(
\frac{2\pi x_0}{w_0}
\right)
\,
e^{-\frac{1}{\alpha}\left(\frac{\pi}{w_0}\right)^2}
\right]
\label{app.eq.Evar.explicit}
\end{eqnarray}
where we let the mass $m$ be the annealing parameter ($s=1$). The conditions for stationary $E^{\mathrm{var}}(\alpha,x_0)$ are
\begin{equation}
 \frac{\partial E^{\mathrm{var}}(\alpha,x_0) }{\partial \alpha}=
\frac{\partial E^{\mathrm{var}}(\alpha,x_0) }{\partial x_0}=0
\label{}
\end{equation}
where the derivatives are 
\begin{equation}
\frac{\partial E^{\mathrm{var}} }{\partial \alpha}
=
\frac{\hbar^2}{4m}
-
\frac{k}{4\alpha^2}
-
\frac{h_0}{2}
\left(
\frac{\pi}{\alpha w_0}
\right)^2
\cos
\left(
\frac{2\pi x_0}{w_0}
\right)
\,
e^{-\frac{1}{\alpha}\left(\frac{\pi}{w_0}\right)^2}
\label{eq.dEv/da}
\end{equation}
and 
\begin{equation}
\frac{\partial E^{\mathrm{var}}}{\partial x_0}
=
k x_0
+
\frac{h_0}{2}
\left(
\frac{2\pi}{w_0}
\right)
\sin
\left(
\frac{2\pi x_0}{w_0}
\right)
\,
e^{-\frac{1}{\alpha}\left(\frac{\pi}{w_0}\right)^2}
\label{eq.dEv/dx0}
\end{equation}

\begin{figure*}[t]
\begin{center}
\includegraphics[scale=0.4]{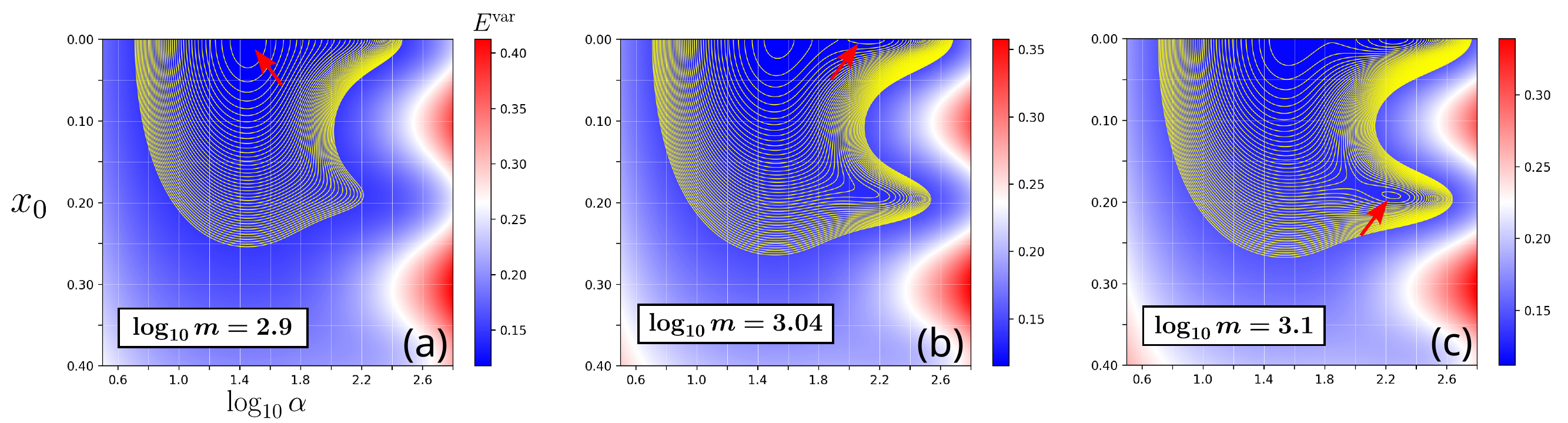}
\caption{Graphs of variational energy $E^{\mathrm{var}}(\alpha,x_0)$ of the Rastrigin system [Eq. (\ref{app.eq.Evar.explicit})], plotted as a heat map (for $h_0=w_0=0.2$ and $\hbar=k=1$). Contour lines of constant $E^{\mathrm{var}}$ (yellow curves) are overlaid to aid visualization. (Along vertical axis, $x_0$ increases downwards.) Arrows (red) indicate minima (i.e., solutions) on the variational surface. (a) $\log_{10} m=2.9$. First solution at $x_0=0$. (b) $\log m=3.04$. Second solution at $x_0=0$. (c) $\log m=3.1$. A third solution at $x_0\approx 0.2$, which is the $x$-coordinate of the lowest local minima of the Rastrigin function.}
\label{fig.app.Rastrigin contour.variational energy}
\end{center}
\end{figure*}

To illustrate, Fig. \ref{fig.app.Rastrigin contour.variational energy} shows the graph of Eq. (\ref{app.eq.Evar.explicit}) (for $h_0=w_0=0.2$ and $\hbar=k=1$, following the parameters in Ref. \cite{Koh22}). It is plotted as a two-dimensional heat map with $x_0$ and $\alpha$ along the vertical and horizontal axes, respectively. To aid visualization, in the region of interest contour lines of constant $E^{\mathrm{var}}$ have been overlaid (yellow curves). Arrows (red) indicate minima on the $E^{\mathrm{var}}(\alpha,x_0)$ surface, solutions of Eqs. (\ref{eq.dEv/da}) and (\ref{eq.dEv/dx0}).

Panel (a) shows the case of $\log_{10} m=2.9$. There is a minimum at $x_0=0$, representing the ground state solution. As $m$ increases, a second minimum appears, also at $x_0=0$. Panel (b) shows the situation at $\log m=3.04$. This new solution initially has higher energy, and so is not physically relevant.  As the mass increases further to $\log m\approx 3.07$, the second solution at $x_0=0$ undergoes a first-order transition to become the ground state; simultaneously, a third solution appears at $x_0\approx 0.2$. Panel (c) shows the surface with the three minima at $\log m=3.1$. The third solution is positioned at the lowest local minimum of the Rastrigin function, and represents the first-excited state. Within the variational framework, the energy gap $E_1-E_0$ is given by the difference between the variational energies of the second and third minima.

This summarizes the general behavior of the $E^{\mathrm{var}}(\alpha,x_0)$ surface with increase in mass. We shall not delve into further details, but simply note that both the ground state energy and the flat gap are numerically very well approximated by the variational approach. In the following, we will focus on understanding how the flatness of the energy gap can be understood within the variational framework.


\subsection{Nature of the gap's flatness}
\label{app.subsec.nature of flat gap}

As we wish to understand why the energy gap exhibits flatness, let us consider the gradient of the variational energy with respect to mass
\begin{eqnarray}
&& \frac{\partial E^{\mathrm{var}}(\alpha,x_0)}{\partial m}
=
-\frac{\hbar^2}{4m^2}
\,
\alpha
+
\frac{\hbar^2}{4m}
\,
\frac{\partial \alpha}{\partial m}
+
\frac{k}{2}
\left[
2x_0
\frac{\partial x_0}{\partial m}  \right. \nonumber \\
&& -\left.
\frac{1}{2\alpha^2}
\frac{\partial \alpha}{\partial m}
\right]
+
\frac{h_0}{2}
\,
e^{-\frac{1}{\alpha}\left(\frac{\pi}{w_0}\right)^2}
\left[
\left(\frac{2\pi}{w_0}\right)
\,
\frac{\partial x_0}{\partial m}
\,
\sin\left(\frac{2\pi x_0}{w_0}\right) \right. \nonumber \\
&& \left.
-\frac{1}{\alpha^2}
\,
\frac{\partial \alpha}{\partial m}
\,
\left(\frac{\pi}{w_0}\right)^2
\,
\cos\left(\frac{2\pi x_0}{w_0}\right)
\right]
\label{app.eq.dEvdm.first expression}
\end{eqnarray}
From Eq. (\ref{eq.dEv/da}), the stationary condition yields
\begin{equation}
\frac{h_0}{2}
\,
\frac{1}{\alpha^2}
\,
\left(\frac{\pi}{w_0}\right)^2
\,
\cos\left(\frac{2\pi x_0}{w_0}\right)
\,
e^{-\frac{1}{\alpha}\left(\frac{\pi}{w_0}\right)^2}
=
\frac{\hbar^2}{4m}
-
\frac{k}{4\alpha^2}
\label{app.eq.rearrange.dEv/da}
\end{equation}
From Eq. (\ref{eq.dEv/dx0}) one has
\begin{equation}
\left(\frac{2\pi}{w_0}\right)
\sin\left(\frac{2\pi x_0}{w_0}\right)
e^{-\frac{1}{\alpha}\left(\frac{\pi}{w_0}\right)^2}
=
-2k
\left(
\frac{x_0}{h_0}
\right)
\label{eq.rearrange.dEv/dx0}
\end{equation}
Using Eqs. (\ref{app.eq.rearrange.dEv/da}) and (\ref{eq.rearrange.dEv/dx0}) to eliminate the sine and cosine terms in Eq. (\ref{app.eq.dEvdm.first expression}), we arrive at the simple result
\begin{equation}
\left.
\frac{\partial \tilde{E}^{\mathrm{var}}}{\partial m}\right
|_{\tilde{\alpha},\tilde{x_0}}
=
-
\frac{\hbar^2}{4m^2}
\,
\tilde{\alpha}
\label{app.eq.dEv/dm.final expression}
\end{equation}
where $(\tilde{\alpha},\tilde{x}_0)$ denotes the solution of Eqs. (\ref{app.eq.rearrange.dEv/da}) and (\ref{eq.rearrange.dEv/dx0}). 

Defining the variational energy gap
\begin{equation}
\Delta^{\mathrm{var}}=
\tilde{E}^{\mathrm{var}}(1)
-
\tilde{E}^{\mathrm{var}}(0)
\label{app.eq.variational gap. definition}
\end{equation}
where $\tilde{E}^{\mathrm{var}}(1)$ and $\tilde{E}^{\mathrm{var}}(0)$ are evaluated at the third and second solutions on the variational surface, respectively (see Fig. \ref{fig.app.Rastrigin contour.variational energy}). Using Eq. (\ref{app.eq.dEv/dm.final expression}), we arrive at our central result
\begin{equation}
\frac{\partial\Delta^{\mathrm{var}}}{\partial m}
=
-\frac{\hbar^2}{4m^2}
\left[
\tilde{\alpha}(1)
-
\tilde{\alpha}(0)
\right]
\label{app.eq.Deltav.final expression}
\end{equation}
where $\tilde{\alpha}(1)$ and $\tilde{\alpha}(0)$ are the third and second solutions, respectively. The reason for the energy gap being `flat' is now clear. The gap gradient is, strictly speaking, non-zero, and depends on two factors. As $m^2$ appears in the denominator, if the mass is large, the gap gradient will be small. For instance, one sees in Fig. \ref{fig.app.Rastrigin contour.variational energy}(c) that $m=10^{3.1}$. 

The second factor is the difference $\tilde{\alpha}(1)-\tilde{\alpha}(0)$. That $\tilde{\alpha}(1)\approx\tilde{\alpha}(0)$ can be seen as follows. For $\tilde{\alpha}(0)$, substituting $x_0=0$ into Eq. (\ref{app.eq.rearrange.dEv/da}), we see that it must satisfy 
\begin{equation}
\frac{h_0}{2}
\,
\frac{1}{\alpha^2}
\,
\left(\frac{\pi}{w_0}\right)^2
\,
e^{-\frac{1}{\alpha}\left(\frac{\pi}{w_0}\right)^2}
=
\frac{\hbar^2}{4m}
-
\frac{k}{4\alpha^2}
\label{app.eq.dEv/da=0.no more x0}
\end{equation}
which is independent of $x_0$. For the solution at the local minima, from Eq. (\ref{eq.rearrange.dEv/dx0}) one has $\tilde{x}_0(1)=w_0+y$, where
\begin{equation}
y
=
\frac{-k w_0}{k+\frac{h_0}{2} \left(\frac{2\pi}{w_0}\right)^2  e^{-\frac{1}{\alpha}\left(\frac{\pi}{w_0}\right)^2}}
\label{app.eq.y soln.when y small}
\end{equation}
is a small shift. This gives $\cos\left(\frac{2\pi\tilde{x}_0(1)}{w_0}\right)\approx 1$, which implies that $\tilde{\alpha}(1)$ again satisfies Eq. (\ref{app.eq.dEv/da=0.no more x0}). This substantiates our assertion that $\tilde{\alpha}(1)\approx\tilde{\alpha}(0)$.


\subsection{General validity}
\label{app.subsec.General validity}

In arriving at the above conclusions, the key step is Eq. (\ref{app.eq.dEv/dm.final expression}). This term originates from the kinetic energy operator in the Hamiltonian. Although we did not furnish a proof here, terms from the potential energy cancel out in the steps leading from Eq. (\ref{app.eq.dEvdm.first expression}) to (\ref{app.eq.dEv/dm.final expression}). In other words, Eq. (\ref{app.eq.dEv/dm.final expression}) requires only the kinetic term $\frac{p^2}{2m}$ and the gaussian ansatz Eq. (\ref{eq.app.trial psi.gaussian}), and does not depend on specific potential energy. If we alter the boundary conditions, as in the case of the box model, the analytic form of the energy gradient $\frac{\partial \tilde{E}^{\mathrm{var}}}{\partial m}$ would certainly change to reflect boundary effects. However, if the global and local minima involved are far from the boundaries, these effects should only be minor perturbations, with Eq. (\ref{app.eq.dEv/dm.final expression}) again being the leading contribution. Hence, for any potential in general, we think that flat gaps can be explained by Eq. (\ref{app.eq.Deltav.final expression}). The role played by $m^2$ is numerically more important, and also more universal because it stems from the kinetic energy. On the other hand, the second factor $\tilde{\alpha}(1)-\tilde{\alpha}(0)$ depends on the potential surface. If the local curvatures of the global and local minima are very similar (which is the case for the Rastrigin function and our box model), the widths of the gaussians localized at these minima would also be similar, and $\tilde{\alpha}(1)\approx\tilde{\alpha}(0)$ would hold. If the curvatures are very different, then the second factor would not play such an important role in the gap's flatness.

%
%


\begin{thebibliography}{99}
\bibitem{notes.Scipy.solveivp} To perform the numerical integration, we used the solve\_ivp function from the SciPy library. The parameter `method' is set to `DOP853', and let both `rtol' and `atol' be either $10^{-10}$, $10^{-11}$, or $10^{-12}$, depending on the precision needed for the final energy.
\end{thebibliography}
\end{document}